\documentclass[
reprint,
superscriptaddress,
nofootinbib,
aps, pra
]{revtex4-1}

\usepackage{custom}
\usepackage{comment}
\usepackage{float}
\usepackage{graphicx} 
\usepackage[caption=false]{subfig}
\begin{document}
\title{Gravitational waves  affect vacuum entanglement}

\author{Qidong Xu}
\email[]{qidong.xu.gr@dartmouth.edu}
\affiliation{Department of Physics and Astronomy, Dartmouth College, Hanover, New Hampshire 03755, USA}

\author{Shadi Ali Ahmad}
\email[]{shadi.ali.ahmad.22@dartmouth.edu}
\affiliation{Department of Physics and Astronomy, Dartmouth College, Hanover, New Hampshire 03755, USA}

\author{Alexander R. H. Smith}
\email[]{alexander.r.smith@dartmouth.edu}
\affiliation{Department of Physics and Astronomy, Dartmouth College, Hanover, New Hampshire 03755, USA}

\date{\today}

\begin{abstract}
The entanglement harvesting protocol is an operational way to probe vacuum entanglement. This protocol relies on two atoms, modelled by Unruh-DeWitt detectors, that are initially unentangled. These atoms then interact locally with the field and become entangled. If the atoms remain spacelike separated, any entanglement between them is a result of entanglement that is `harvested' from the field. Thus, quantifying this entanglement serves as a proxy for how entangled the field is across the regions in which the atoms interacted. Using this protocol, it is demonstrated that while the transition probability of an individual inertial atom is unaffected by the presence of a gravitational wave, the entanglement harvested  by two atoms depends sensitively on the frequency of the gravitational wave, exhibiting novel resonance effects when the energy gap of the detectors is tuned to the frequency of the gravitational wave. This suggests that the entanglement signature left by a gravitational wave may be useful in characterizing its properties, and potentially useful in exploring the gravitational-wave memory effect and gravitational-wave induced decoherence.  
\end{abstract}

\maketitle

\section{Introduction}
\label{Introduction}

 It has long been realized that the vacuum state of a quantum field theory in Minkowski space is highly entangled across spacelike regions; for example see~\cite{witten2018aps} and references therein.  Using algebraic methods, Summers and Werner demonstrated that correlations between field observables across spacelike regions are strong enough to violate a Bell inequality~\cite{summersBellInequalitiesQuantum1987,summersBellInequalitiesQuantum1987a,summersVacuumViolatesBell1985a}. It was later realized that this vacuum entanglement could be `harvested' by atoms\,/\,detectors that couple locally to the field~\cite{valentiniNonlocalCorrelationsQuantum1991,reznikEntanglementVacuum2003,reznikViolatingBellInequalities2005}. This result is surprising, suggesting that the vacuum is a resource for quantum correlations and has since been examined in a wide range of \mbox{scenarios~\cite{martin-martinezSustainableEntanglementProduction2013,saltonAccelerationassistedEntanglementHarvesting2015a,ralphQuantumKeyDistribution2015,pozas2015harvesting,pozas-kerstjensEntanglementHarvestingElectromagnetic2016,martin-martinezPreciseSpaceTime2016,sachsEntanglementHarvestingDivergences2017,ardenghiEntanglementHarvestingDoublelayer2018,trevisonPureStateEntanglement2018,martin-martinezRelativisticQuantumOptics2018,simidzijaHarvestingCorrelationsThermal2018,congEntanglementHarvestingMoving2019,hendersonBandlimitedEntanglementHarvesting2020,hendersonQuantumTemporalSuperposition2020,faureParticleDetectorsWitnesses2020a}.}

This phenomenon can be used to construct an operational measure of vacuum entanglement. Specifically, supposing that two detectors remain spacelike separated for the duration of their interaction with the field, then any entanglement that results between them must be attributed to entanglement `harvested' from the vacuum that existed prior to the detectors' interaction. Thus, quantifying how entangled two detectors become serves as a proxy for how entangled the vacuum is across the regions in which the detectors have interacted. Such a quantification of vacuum entanglement is similar to the distillable entanglement defined as the number of maximally entangled states that can be `distilled' from a number of copies of a given quantum state via local operations and classical communication~\cite{plenioIntroductionEntanglementMeasures2007}.

Entanglement harvesting has been used to probe the effects of nontrivial spacetime structure on vacuum entanglement, such as cosmological effects~\cite{steegEntanglingPowerExpanding2009,martin-martinezCosmologicalQuantumEntanglement2012,martin-martinezSustainableEntanglementProduction2013,martin-martinezEntanglementCurvedSpacetimes2014,huangDynamicsQuantumEntanglement2017}, nontrivial spacetime topology~\cite{martin-martinezSpacetimeStructureVacuum2016, lin2016entanglement, smithDetectorsReferenceFrames2019}, spacetime curvature~\cite{clicheVacuumEntanglementEnhancement2011,ngUnruhDeWittDetectorsEntanglement2018,ngNewTechniquesEntanglement2018a,hendersonEntanglingDetectorsAntide2019}, and black hole  horizons~\cite{hendersonHarvestingEntanglementBlack2018,congEffectsHorizonsEntanglement2020}. It is the purpose of this article to extend this analysis to examine how a gravitational wave affects the entanglement structure of the vacuum. To do so, we derive the gravitational wave modification to the Minkowski space Wightman function and evaluate the final state of two detectors that are initially unentangled. The final state of the detectors is entangled, and the amount of entanglement depends sensitively on the frequency of the gravitational wave and detectors' energy gap. In particular, we demonstrate that a resonance effect occurs when the detectors' energy gap is tuned to the frequency of the gravitational wave. If the detectors' interaction is centered around the gravitational wave's peak displacement, then the gravitational wave is shown to degrade the harvested entanglement relative to detectors in Minkowski space. However, when the detectors' interaction is not centered at this point in the gravitational wave's cycle, then the harvested entanglement can be either amplified or degraded and oscillates as a function of gravitational wave frequency. Away from this resonance condition, the effect of a gravitational wave on the harvested entanglement is exponentially suppressed. 

Moreover, we demonstrate that the transition probability  of an inertial detector is unaffected by the presence of a gravitational wave, and thus does not register a different particle content than if it were in Minkowski space. This is consistent with Gibbons' conclusion that  gravitational waves do not produce particles~\cite{gibbonsQuantizedFieldsPropagating1975}. In contrast, we emphasize that the entanglement  between two detectors is sensitive to the presence of a gravitational wave. This result is analogous to the observation made by ver Steeg and Menicucci~\cite{steegEntanglingPowerExpanding2009} that a single detector is unable to distinguish the field being in a thermal state in Minkowski space or the vacuum in a de Sitter spacetime, whereas the correlations between two detectors can distinguish between these situations. Furthermore, this result agrees with the intuition from the classical theory of gravitational waves which asserts that a gravitational wave cannot be detected by a local detector moving along a geodesic.

\section{Scalar field theory in a gravitational wave background}
\label{deriveWightmanfunc}

A gravitational wave propagating along the $z$-direction is described by the line element
\begin{align}
ds^2 &= -dt^2 +dz^2 +(1+A \cos\left[\omega (t-z)\right]) dx^2 \label{LineElement}  \\ 
&\quad + (1-A \cos\left[ \omega(t-z)\right])dy^2 \nn \\
& =- du dv +(1+ A \cos \omega u ) dx^2  + (1-A \cos \omega u)dy^2, \nn
\end{align}
where in the last equality we have introduced light cone coordinates $u \ce t-z$ and $v \ce t+z$ defined in terms of  Minkowski coordinates $(t,x,y,z)$. On this spacetime, consider a massless scalar field $\phi(\mathsf{x})$ satisfying the Klein-Gordon equation at a spacetime point $\mathsf{x}$,
\begin{align}
\Box  \phi(\mathsf{x}) = 0,
\label{KGequation}
\end{align}
where $\Box$ is the d'Alembertian operator associated with Eq.~\eqref{LineElement}.\footnote{We could have considered a nonminimal coupling of the field to the Ricci scalar by including a term $\xi R$ in the equation above. However, for a gravitational wave spacetime like the one described in Eq.~\eqref{LineElement} $R$ vanishes.}  Solving this equation in light-cone coordinates $\mathsf{x} = (u,v,x,y)$ yields a complete set of  solutions~\cite{garrigaScatteringQuantumParticles1991}
\begin{align}
u_{\vec{k}}(\mathsf{x}) =& \frac{\gamma^{-1}(u )}{\sqrt{2k_{-}}(2\pi)^{\frac{3}{2}}} e^{ik_a x^a -ik_{-}v  -\frac{i}{4k_{-}}\int_0^u du \,  (g^{ab}k_a k_b )},
\label{modelfunction} 
\end{align} 
where $\gamma^{-1}(u) \ce [\det g_{ab}(u)]^{\frac{1}{4}}$, the indices $a$ and $b$ run over $\{x, y\}$, and $\vec{k} \ce (k_{-}, k_{a})$  are separability constants arising from solving Eq.~\eqref{KGequation}  in light-cone coordinates.  This set of solutions is orthonormal with respect to the usual Klein-Gordon inner product \cite{garrigaScatteringQuantumParticles1991,BirrellDavies}.

Quantization proceeds by promoting the field to an operator and imposing the canonical commutation relations~\cite{BirrellDavies,Wald}. As the solutions to Eq.~\eqref{KGequation} are most easily constructed in light cone coordinates, we quantize the field in this coordinate system. For a free field theory, light cone quantization has been shown to be equivalent to the more familiar equal time quantization procedure~\cite{mannheim2020light}.  Thus, we can interpret the mode functions in Eq.~\eqref{modelfunction} as describing the perturbation to the Minkowski vacuum induced by a gravitational wave. As we shall see, using light cone quantization yields the same detector behaviour in the Minkiwoski space limit ($A\to 0$) as equal-time quantization.

As derived in Appendix~\ref{Derivation of Wightman function}, the vacuum Wightman function is
\begin{align}
W(\mathsf{x},\mathsf{x}') &\ce \braket{0|\phi(\mathsf{x}) \phi(\mathsf{x}') | 0} = \int d \mathsf{k} \, u_{\mathsf{k}}(\mathsf{x}) u_{\mathsf{k}}^*(\mathsf{x}') \nn \\
&= W_{ \mathcal{M}}(\mathsf{x},\mathsf{x}') + W_{\rm GW}(\mathsf{x},\mathsf{x}'), \notag
\end{align}
where $W_{ \mathcal{M}}(\mathsf{x},\mathsf{x}')$ is the Minkowski space Wightman function which is independent of the gravitational wave in light-cone coordinates,
\begin{align}
W_{\mathcal{M}} (\mathsf{x},\mathsf{x}') &=\! \frac{1}{4\pi i \Delta u  }    \delta\left(\frac{\sigma_{\mathcal{M}}(\mathsf{x},\mathsf{x}')}{\Delta u}  \right)  
+  \frac{1}{4\pi^2  \sigma_{\mathcal{M}}(\mathsf{x},\mathsf{x}')} ,
\nn
\end{align}
where  $\Delta \mathsf{x}^\mu \ce \mathsf{x}^\mu-\mathsf{x}'^\mu$, and
\begin{align}
\sigma_{\mathcal{M}}(\mathsf{x},\mathsf{x}') \ce - \Delta u\Delta v+  \Delta x^2  +  \Delta y^2, \nn 
\end{align}
 is the geodesic distance between $\mathsf{x}$ and $\mathsf{x}'$ in Minkowski space, and the modification of the Minkowski Wightman function to first order in the gravitational wave amplitude 
 \begin{align}
W_{\rm GW}(\mathsf{x},\mathsf{x}') &=\! -\tfrac{A}{4 \pi^{2}  }  \sinc \left( \tfrac{ \omega}{2} \Delta u  \right)  \cos  \left( \tfrac{\omega}{2}    [u+u'] \right)   \nn \\ 
&\! \quad \times\! \tfrac{ \Delta x^2 - \Delta y ^2  }{\Delta u^2}\!  
 \left[ i \pi \delta'\!\left(  \tfrac{\sigma_{\rm \mathcal{M}}(\mathsf{x},\mathsf{x}')}{\Delta u} \right)  \! + \!\tfrac{\Delta u ^2}{\sigma_{\mathcal{M}}^2(\mathsf{x},\mathsf{x}') }  \right]\!,
\label{WGW}
\end{align}
where $\sinc{x} \ce \frac{\sin x}{x}$.

\section{Detectors in the presence of gravitational waves}
\label{mainresult}

To operationally probe the effects a gravitational wave has on the vacuum state of a scalar field theory, we employ so-called Unruh-DeWitt detectors. Such detectors are a model of a two-level atom locally coupled to a quantum field. We use these detectors to probe interesting field observables in a gravitational wave background, and to track their deviation from the equivalent observables  in Minkowski space. After describing these detectors in detail, we demonstrate that the transition probability of an inertial detector is unaffected by the presence of a gravitational wave.

Then, two initially uncorrelated detectors will be used to examine the effect a gravitational wave has on vacuum entanglement by quantifying how entangled they become as a result of their interaction; this protocol will be referred to as \emph{entanglement harvesting}. We demonstrate that the entanglement harvested by the detectors depends sensitively on the gravitational wave frequency $\omega$ and exhibits resonance effects.

\subsection{The Unruh-DeWitt detectors and the light-matter interaction}
The Unruh-DeWitt detector~\cite{unruhNotesBlackholeEvaporation1976,dewittGeneralRelativityEinsten1979} is a simplified model of a two-level atom, with a ground state  $\ket{0_D}$ and excited state $\ket{1_D}$, separated by an energy gap $2\Omega$. The center of mass of the detector  is taken to move along the classical spacetime trajectory $\mathsf{x}_D(t)$ parametrized by the detector's proper time $t$. As an approximation to the light-matter  interaction, the detector couples locally with the scalar field $\phi(\mathsf{x})$ along its trajectory.  In the interaction picture, the Hamiltonian describing this interaction~is
\begin{align}
H_D(t) &= \lambda \chi\! \left(t \right)\Big(e^{ i\Omega t} \sigma^+  +  e^{- i\Omega t}\sigma^- \Big) \otimes  \phi\left[\mathsf{x}_D(t)\right], \label{InteractionHamiltonian}
\end{align}
where $\lambda$ is the strength of the interaction,  $\chi(t) \ce e^{-\frac{(t-t_0)^2}{2\sigma^2} } $ is a switching function with the interpretation that $t_0$ and $\sigma$ correspond to when the interaction takes place and its duration, respectively, and $\sigma^+ \ce  \ket{1_D}\!\bra{0_D}$ and $\sigma^- \ce  \ket{0_D}\!\bra{1_D}$ are ladder operators acting on the detector Hilbert space.  Although simple, this model captures the relevant features of the light-matter interaction when no angular momentum exchange is involved~\mbox{\cite{martin-martinezWavepacketDetectionUnruhDeWitt2013,alhambraCasimirForcesAtoms2014,pozas-kerstjensEntanglementHarvestingElectromagnetic2016,martin-martinezRelativisticQuantumOptics2018}. }

\subsection{Single detector excitation as a proxy for vacuum fluctuations}

If an Unruh-DeWitt detector begins ($t\to - \infty$) in its ground state $\ket{0_D}$, due to fluctuations of the vacuum and a finite interaction time, there is a finite probability $P$ that in the far future ($t \to \infty$) it will transition to its excited state $\ket{1_D}$. The probability of such a transition is given to leading order in the interaction strength by~\cite{Loukotransition2007,Loukoclick2006}
\begin{align}
\label{tprob}
P &= \lambda^2 \int_{-\infty}^\infty dt  d t' \, \chi(t) \chi(t') e^{-i \Omega \left(t-t'\right)} W\!\left(\mathsf{x}_D(t) , \mathsf{x}_D(t')\right). 
\end{align}
This probability may be interpreted as quantifying the ability of a detector (or atom) to be spontaneously excited by vacuum fluctuations. Suppose that the detector is at rest with respect to the Minkowski coordinates introduced in Eq.~\eqref{LineElement}, so that its trajectory is the geodesic
\begin{align}
\mathsf{x}_D(t) &= (t, 0, 0, 0).
\label{SingleTrajectory}
\end{align}
 Note that for this detector trajectory, the gravitational wave contribution to the Wightman function in Eq.~\eqref{WGW} vanishes because $\Delta x^2= \Delta y^2=0$. It follows that the transition probability in Eq. \eqref{tprob} is not affected by the gravitational wave background.  We thus conclude that a single detector cannot detect the presence of a gravitational wave.

The transition probability can be calculated for the trajectory in Eq.~\eqref{SingleTrajectory}, and coincides with the transition probability for a detector in Minkowski space using an equal-time quantization scheme
\begin{align}
   P = \frac{\lambda^2}{4 \pi} \left[ e^{-\sigma^2 \Omega^2} - \sqrt{\pi} \sigma \Omega \left(1- \erf[\sigma \Omega]\right)\right], \nn
\end{align}
see Appendix~\ref{Derivation of P, X and C} for details. The fact that a detector clicks with the same probability as in the Minkowski vacuum is consistent with Gibbons' observation that a gravitational wave will not create particles from the vacuum during its propagation~\cite{gibbonsQuantizedFieldsPropagating1975}.\footnote{This conclusion was arrived at by evaluating the Bogolyubov coefficients between the in and out Minkowski-like regions that sandwich a gravitational wave spacetime and demonstrating the absence of particle creation. This setup models a gravitational wave traveling in Minkowski space. In backgrounds other than Minkowski, gravitational wave perturbations may cause particle production \cite{suBlackHoleSqueezers2017}.}

\subsection{Detector entanglement as a proxy for vacuum entanglement}
\label{detectorentanglementanalysis}

To operationally probe vacuum entanglement across spacetime regions, consider two detectors, $A$ and $B$, each interacting locally with the field $\phi$ for a finite amount of time, after which the detectors become correlated~\cite{valentiniNonlocalCorrelationsQuantum1991, reznikViolatingBellInequalities2005,reznikEntanglementVacuum2003}. If these detectors remain spacelike separated for the duration of their interaction with the field, then any correlations that arise between them must have been harvested from the vacuum state of the field. Thus, their behaviour serves as an operational proxy of vacuum correlations. If it is not the case that the detectors remain spacelike separated, then again correlations may be transferred from the vacuum state of the field to the detectors. However, in this case even though the detectors do not interact directly, they can still be coupled by a field-mediated interaction, that may now have the time to propagate between the detectors leading to detector correlations.

\begin{figure*}[t]
\subfloat[]{
\includegraphics[height = 2.in]{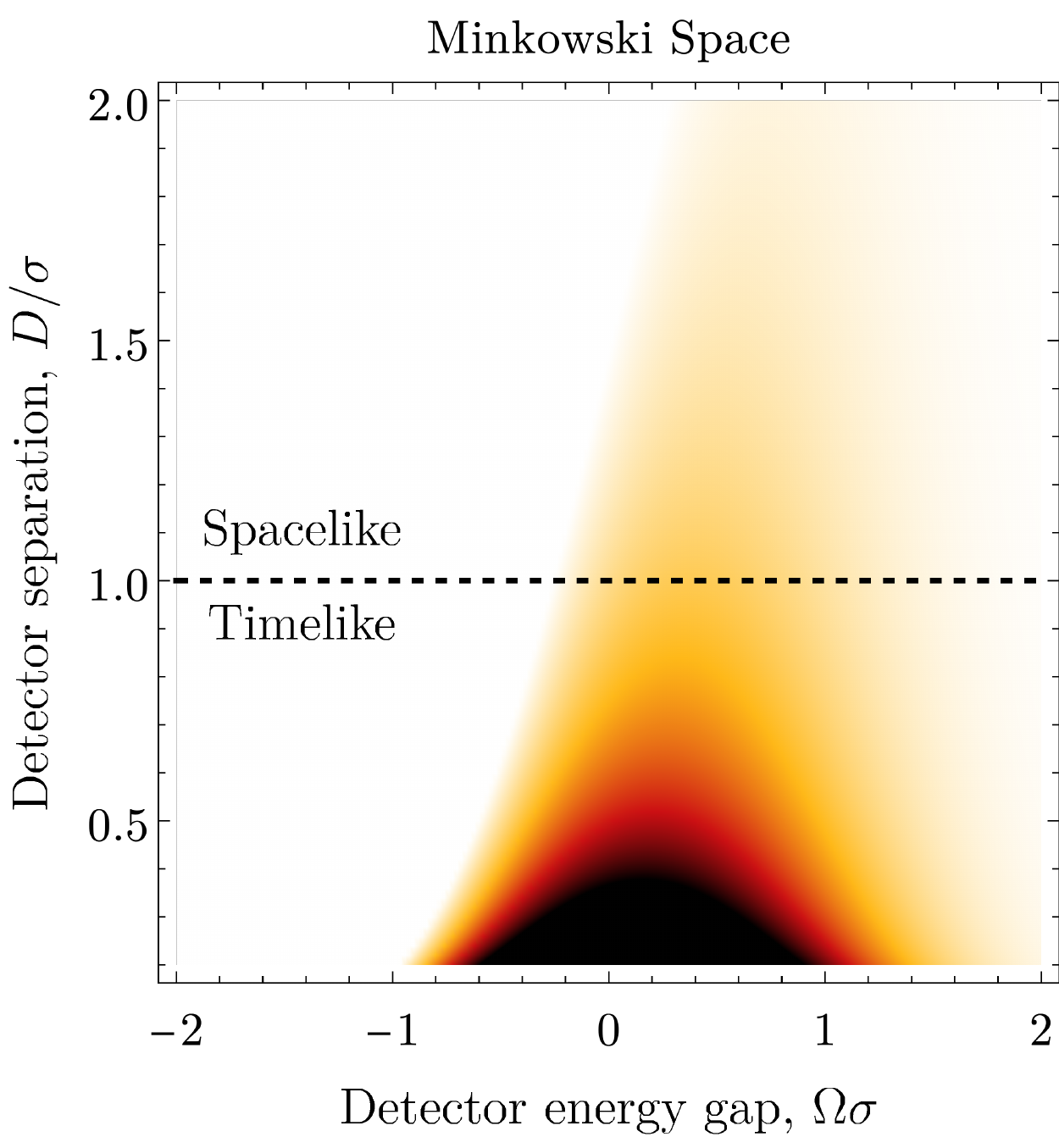} } \quad \  
\subfloat[]{
\includegraphics[height = 2.in]{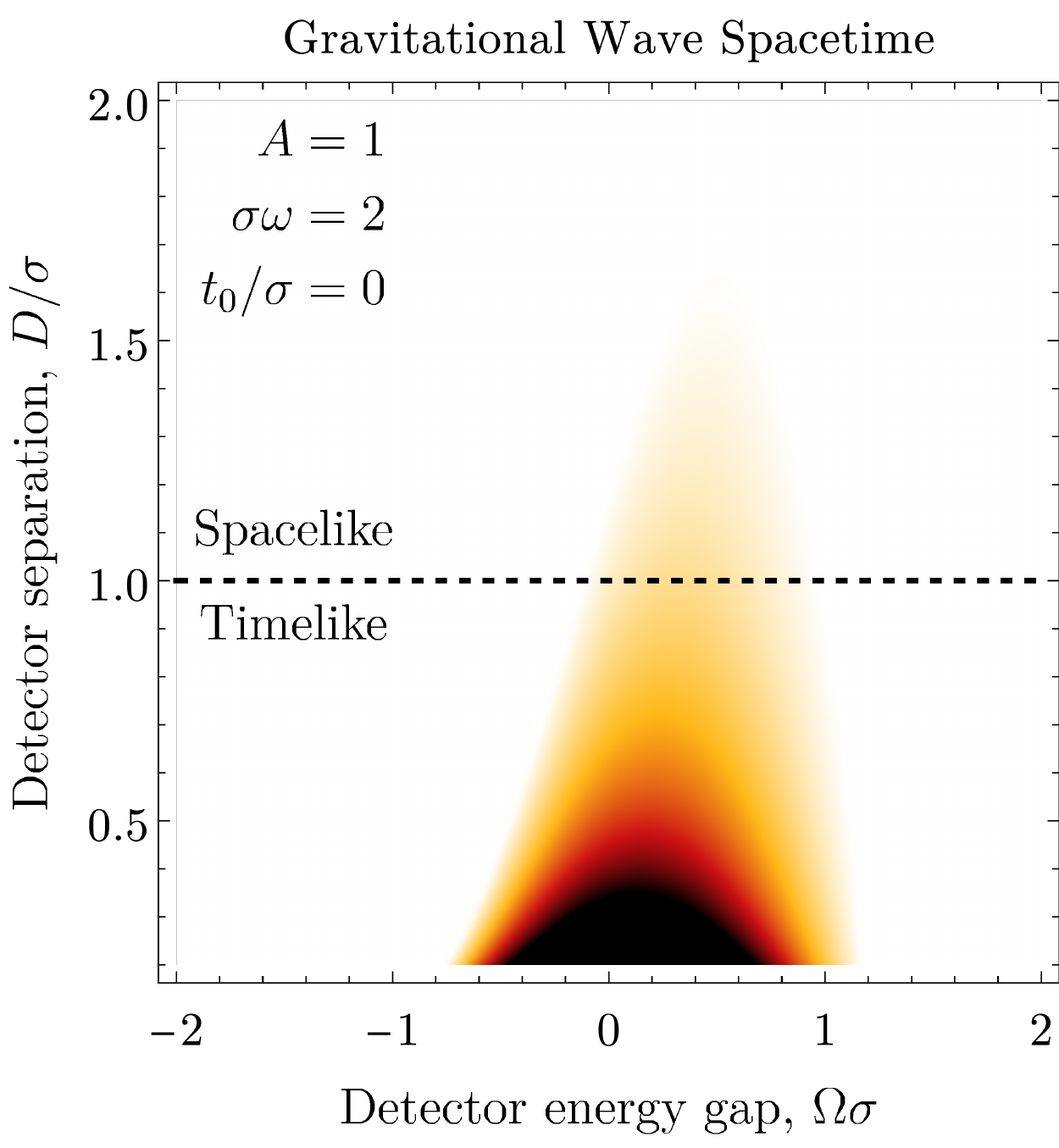} } \quad \
\subfloat[]{
\includegraphics[height = 2.in]{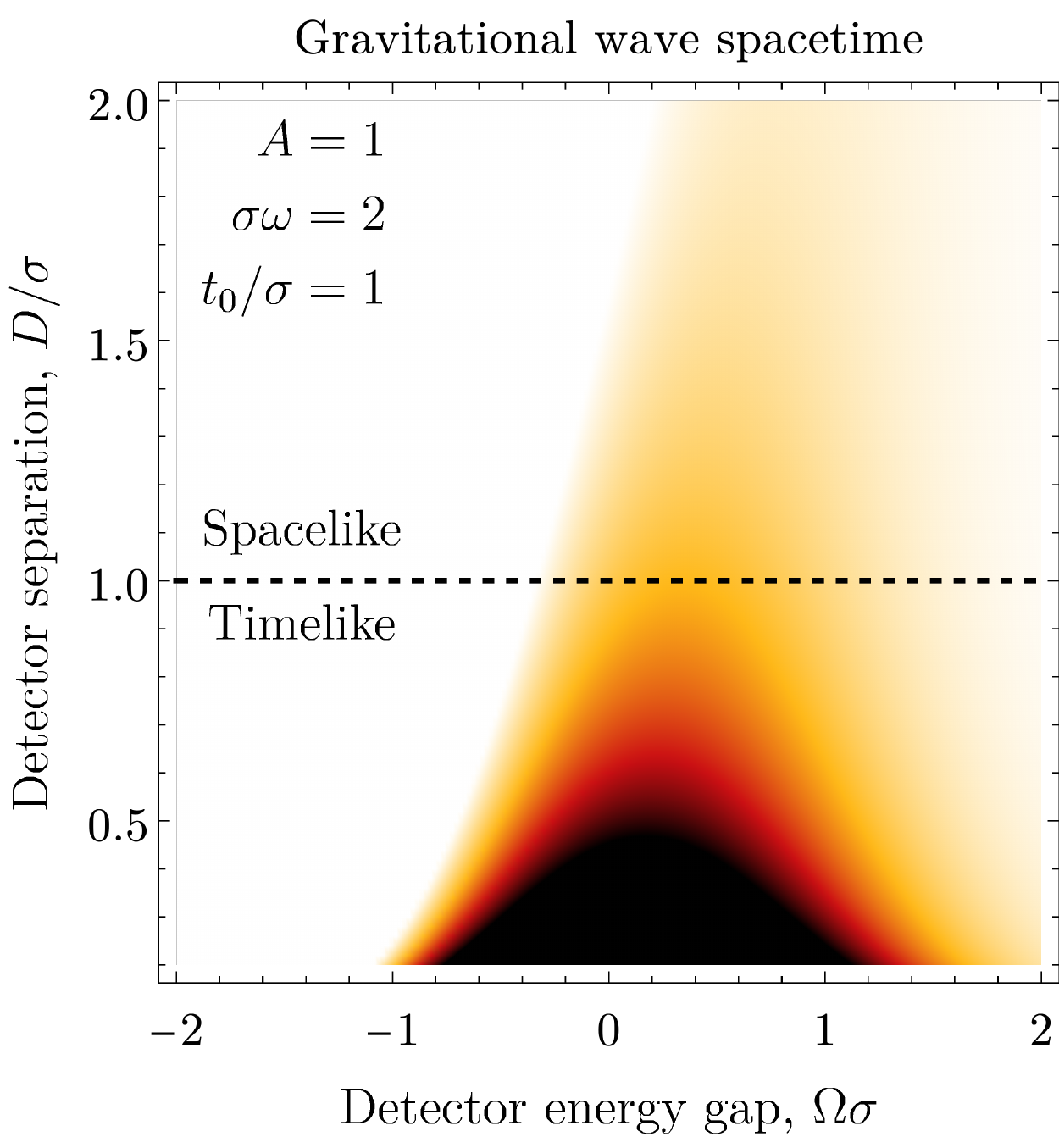} } \quad \
\subfloat{
\includegraphics[height = 2.in]{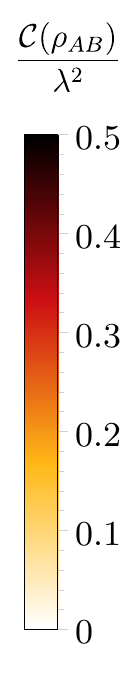} }
\caption{The concurrence $\mathcal{C}(\rho_{AB})/\lambda^2$  is plotted as a measure of entanglement between the two detectors as a function of their energy  $\Omega \sigma$ and average proper separation $D/\sigma$ for detectors situated in (a) Minkowski space  and a gravitational wave spacetime with (b) $t_0=0$  and  (c) with $t_0 = 1$. The gravitational wave contribution degrades the concurrence relative to detectors in Minkowski space for $t_0 = 0$, as can be seen by comparing (a) and (b); however, for $t\neq 0$, as shown in (c),  the concurrence can either be amplified or degraded due to the presence of a gravitational~wave.}
\label{fig:Concurrence}
\end{figure*}

Consider the following trajectories of detectors $A$ and $B$ specified in Minkowski coordinates 
\begin{align}
\mathsf{x}_A(t) &= ( t , 0, 0, 0), \nn \\
\mathsf{x}_B(t) &= ( t, D, 0 , 0).
\label{twodetectortraj}
\end{align}
Note that since the detectors interact with the field for an approximate amount of proper time $\sigma$, detectors moving along these trajectories can be considered approximately spacelike separated when $D > \sigma$; $D$ corresponds to the average proper distance between the detectors. Furthermore, suppose these detectors are initially ($t \to -\infty$) prepared in their ground state, and the state of the field is in an appropriately defined vacuum state $\ket{0}$, so that the joint state of the detectors and field together is $\ket{\Psi_i} = \ket{0}_A \ket{0}_B \ket{0}$. Given that the interaction between each detector and the field is described by the Hamiltonian in Eq.~\eqref{InteractionHamiltonian}, the final ($t \to \infty$) state of the detectors and field is
\begin{align}
\ket{\Psi_f} = \mathcal{T} e^{  -i \int_{\mathbb{R}}  dt\, \left[ H_A( t) +  H_B(t) \right] } \ket{\Psi_i}, \nn
\end{align}
where $H_A$ and $H_B$ are given in Eq.~\eqref{InteractionHamiltonian} and $\mathcal{T}$ denotes the time ordering operator. The reduced state of the detectors is obtained by tracing over the field
\begin{align}
\rho_{AB} &\ce \tr_\phi \big( \ket{\Psi_f}\!\bra{\Psi_f} \big)\nn \\
&= \begin{pmatrix}
1 - 2 P  & 0 & 0 & X \\
0 & P  & C & 0 \\
0 & C^* & P & 0 \\
X^* & 0 & 0 & 0
\end{pmatrix} + \mathcal{O}\!\left(\lambda^4\right), \label{FinsalState2}
\end{align}
expressed in the basis $\{ \ket{0_A 0_B}, \ket{0_A 1_B}, \ket{1_A 0_B}, \ket{1_A 1_B} \}$, and the matrix elements  $X$ and $C$ are given by integrals over the Wightman function evaluated along the detectors' trajectories and are computed analytically in Appendix~\ref{Derivation of P, X and C}.  These matrix elements are the sum of two terms, $X= X_{\mathcal{M}} + X_{\rm GW}$ and $C = C_{\mathcal{M}} + C_{\rm GW}$. The first terms, $X_{\mathcal{M}}$ and $C_{\mathcal{M}}$, correspond to the value $X$ and $C$ would take if the detectors were situated in Minkowski space and coincides with the result obtained using equal-time quantization~\cite{martin-martinezSpacetimeStructureVacuum2016,smithDetectorsReferenceFrames2019},
\begin{align}
X_{\mathcal{M}} \!  &\ce i \frac{\sigma \lambda^2}{4D \sqrt{\pi}} e^{-\sigma^{2}\Omega^2 -2i\Omega t_{0}- \frac{D^{2}}{ 4\sigma^{2}}} \left[ \erf \left(  \frac{iD}{2\sigma}\right) -1 \right]\!, \label{xmexp}\\
C_{\mathcal{M}} &\ce  \frac{\sigma \lambda^2}{4 D\sqrt{\pi}} e^{-\frac{D^{2}}{ 4\sigma^{2}} } \nn \\
&\quad \times \left(\Im \left[ e^{i D \Omega} \erf \left( i \frac{D}{2\sigma} +  \sigma \Omega \right)\right]   - \sin  \Omega D \right) .\nn
\end{align}
The second terms, $X_{\rm GW}$ and $C_{\rm GW}$, correspond to the modification to the matrix elements $X$ and $C$ stemming from the gravitational wave 
\begin{subequations}
\begin{align}
X_{\rm GW} &\ce
\frac{ A \sigma \lambda^2 }{ 4D^2\pi^{3/2}}  f(\omega, \Omega, \sigma,t_0)   \left(I_1 + I_2 \right)  \nn, \\
C_{\rm GW} &\ce -\frac{  A \sigma \lambda^2  }{ 4D^2\pi^{3/2}}    e^{- \frac{ \sigma^{2}\omega^{2}}{4}}  \cos \left(\omega t_{0}\right)  \left( I_3 + I_4 \right),\nn
\end{align} 
\label{expressionforCandX}\end{subequations}
where the terms $I_{1}$ and $I_{2}$ are complicated functions of $\omega$, $D$, and $\sigma$ and the terms $I_{3}$ and $I_{4}$ are complicated functions of $\omega$,  $D$, $\sigma$, and $\Omega$,  which  have been defined in Appendix~\ref{Derivation of P, X and C}, and 
\begin{align}
f(\omega, \Omega, \sigma,t_0) &\ce e^{-\frac{\sigma^2}{4}(\omega - 2\Omega)^2 - it_0(\omega+ 2\Omega)} \nn \\
&\quad +e^{-\frac{\sigma^2}{4}(\omega +2\Omega)^2 + it_0 ( \omega- 2\Omega )}.
\label{functionf}
\end{align}

\begin{figure}[t]
\begin{center}
\includegraphics[height = 2.4in]{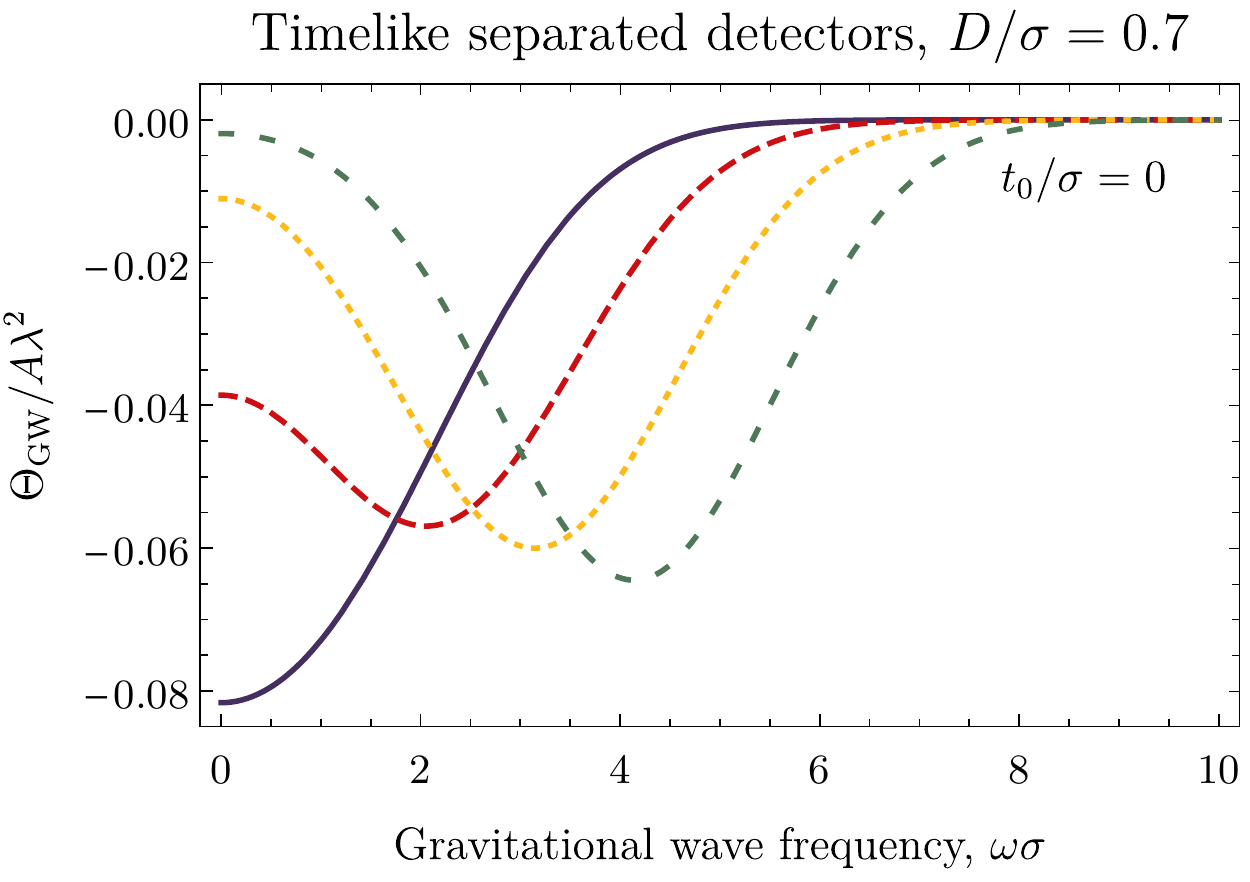} \\ \ \\ 
\includegraphics[height = 2.4in]{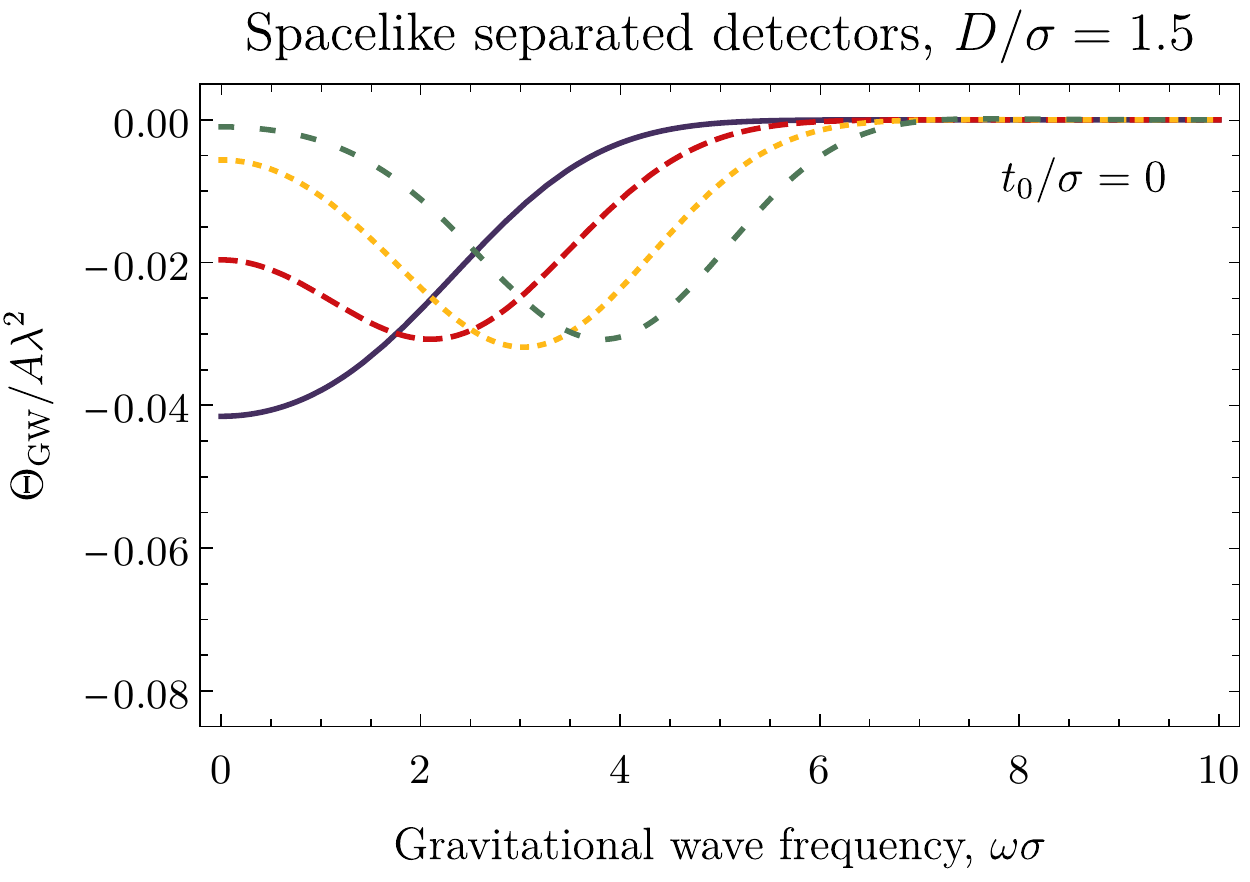} \\
\includegraphics[width = 3.25in]{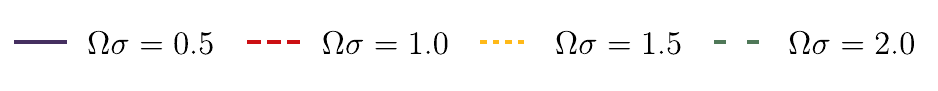} 
\end{center}
\caption{The gravitational wave contribution $\Theta_{\rm GW}/A\lambda^2$  to the concurrence is plotted as a function of the gravitational wave frequency $\omega \sigma$ for both timelike  (\emph{top}) and spacelike  (\emph{bottom}) seperated detectors for $t_0 = 0$. We see that around the resonance condition $\omega \approx 2 \Omega$ the gravitational contribution is negative, which implies a degradation of harvested entanglement relative to detectors in Minkowski space.
\label{fig:omegaVSconcurrence}}
\end{figure}

To quantify the entanglement harvested by the detectors, which will serve as a proxy measure for vacuum entanglement, we use the concurrence as an entanglement measure~\cite{woottersEntanglementFormationConcurrence2001}. For the two detector state in Eq.~\eqref{FinsalState2} the concurrence is~\cite{martin-martinezSpacetimeStructureVacuum2016,smithDetectorsReferenceFrames2019}
\begin{align}
\mathcal{C}(\rho_{AB})= 2 \max \!\left[ \, 0, \ \abs{X} -  P\, \right]  + \mathcal{O}\!\left(\lambda^4\right) 
. \nn
\end{align}
Being a simple difference of a local  term $P$ and non-local term $\abs{X}$, the concurrence $\mathcal{C}(\rho_{AB})$ is convenient in interpreting the results to follow. The concurrence can be expressed as sum of the Minkowski space contribution~$\Theta_{\mathcal{M}}$ and the modification due to the gravitational wave~$\Theta_{\rm GW}$
\begin{align}
\mathcal{C}(\rho_{AB})= 2 \max \!\left[ \, 0, \Theta_{\mathcal{M}} + \Theta_{\rm GW} \right]  + \mathcal{O}\!\left(\lambda^4\right) 
\label{concurrence2},
\end{align}
where
\begin{align}
\Theta_{\mathcal{M}} &\ce \abs{X_{\mathcal{M}}} - P, \nn \\
\Theta_{\rm GW} &\ce   \frac{\Re \left[X_{\rm GW} X_{\mathcal{M}}^{*}\right]}{\abs{X_{\mathcal{M}}}}.
\label{GWcontributiontoconcrrrence}
\end{align}
Note that $\Theta_{\rm GW}$ has been expanded to first order in the gravitational wave amplitude $A$, since this analysis is within the linearized gravity regime.

Figure \ref{fig:Concurrence} compares the behaviour of the concurrence of the final state of two detectors in Minkowski space with an equivalent pair of detectors in the presence of a gravitational wave as a function of the detectors' energy $\Omega \sigma$ and their separation $D/\sigma$; both $t_0 =0$ and $t_0 \neq 0$ are depicted.\footnote{Notice that we choose to survey detector energies $\Omega \sigma \in (-2,2)$. This upper bound is to ensure the validity of the Taylor expansion $A$ to first-order in Eq.~\eqref{concurrence2}. To be more precise, in the numerator of Eq.~\eqref{GWcontributiontoconcrrrence}, $X_{\mathcal{M}}$ approaches zero as $\Omega \sigma$ is getting larger, which causes the second order in $A$ contribution (which would only depend on $X_{\rm GW}$) to dominate~$\Theta_{\rm  GW}$.\label{footnote3}} Since $X_{\mathcal{M}}$ only depends on $t_0$ through an overall phase in Eq.~\eqref{xmexp} and $\Theta_{\rm \mathcal{M}}$ depends on $\abs{X_{\mathcal{M}}}$, the Minkowski contribution to the harvested entanglement is unaffected by $t_0$. From Fig.~\ref{fig:Concurrence}, it is seen that in all instances the concurrence (and thus vacuum entanglement) falls off as the  distance $D/\sigma$ between the detectors grows; this could have been anticipated by noting that both $X_\mathcal{M}$ and $X_{\rm GW}$ are proportional to $e^{-D^2 / 4\sigma^2}$. More interestingly,  Fig.~\ref{fig:Concurrence}b illustrates that for $t_0 =0$ a gravitational wave degrades the concurrence when compared to an equivalent pair of detectors in Minkowski space (Fig.~\ref{fig:Concurrence}a). However, when $t_0 \neq 0$, a gravitational wave can both  amplify or degrade the concurrence depending on the detector separation and gravitational wave frequency, as can be seen in Fig.~\ref{fig:Concurrence}c.

\begin{figure}[t]
\begin{center}
\includegraphics[height = 2.4in]{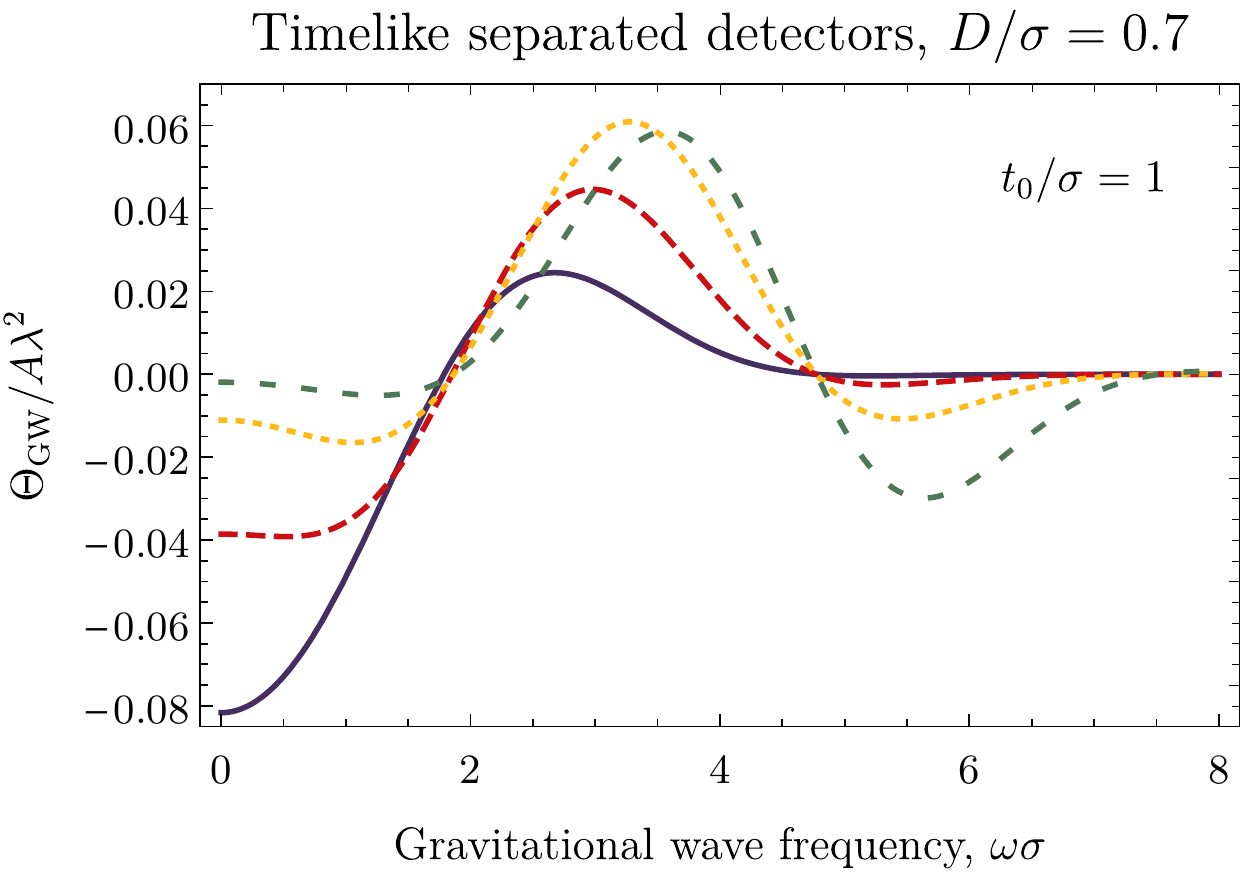} \\ \ \\ 
\includegraphics[height = 2.4in]{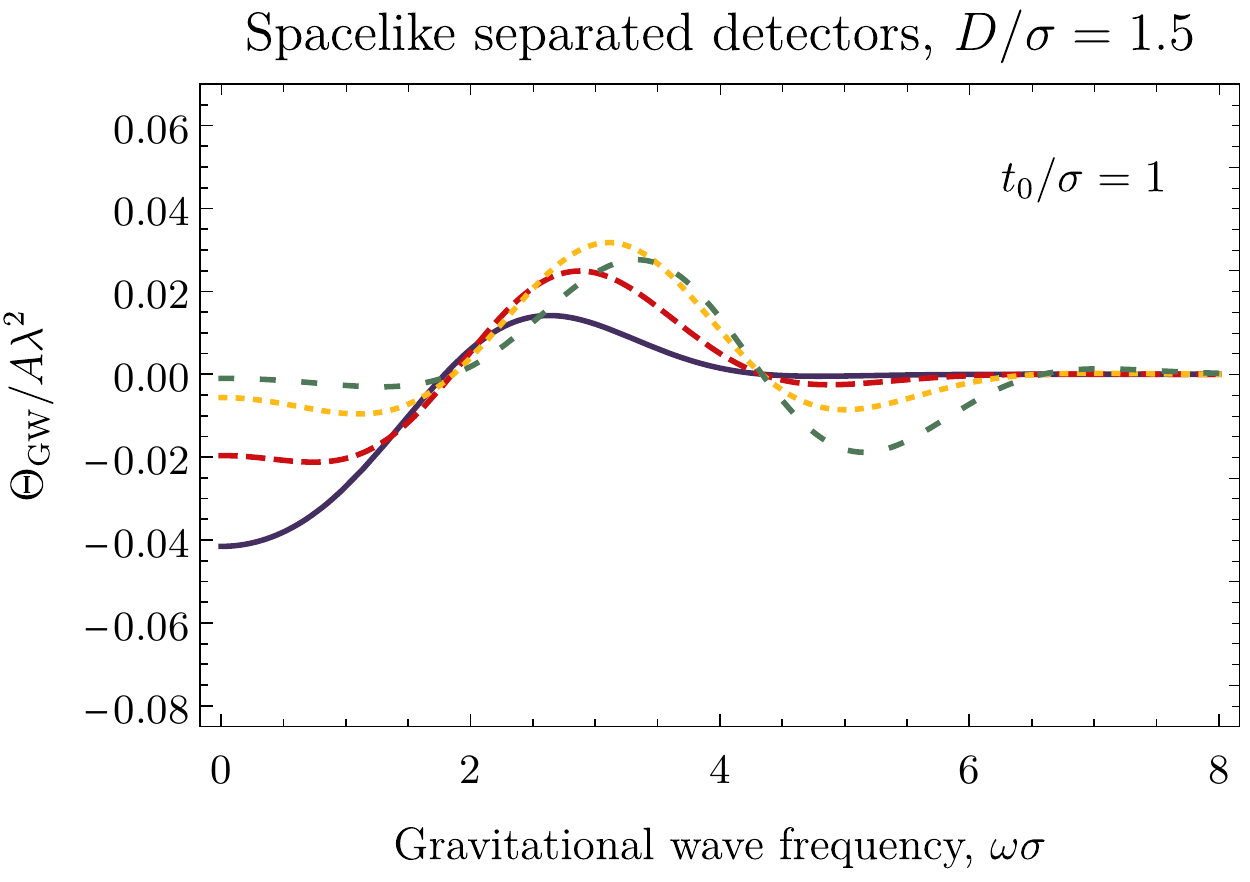} \\
\includegraphics[width = 3.25in]{RowLegend.pdf} 
\end{center}
\caption{The gravitational wave contribution $\Theta_{\rm GW}/A\lambda^2$  to the concurrence is plotted as a function of the gravitational wave frequency $\omega \sigma$ for both timelike  (\emph{top}) and spacelike  (\emph{bottom}) separated detectors for $t_0/\sigma = 1$. We see that around the resonance condition $\omega \approx 2 \Omega$ the gravitational contribution oscillates around zero, which implies that the gravitational wave can either amplify or degrade the harvested entanglement relative to detectors in Minkowski space.}
\label{fig:omegaVSconcurrencet0}
\end{figure}

A more detailed study of the gravitational wave contribution to the concurrence  is shown in Figs.~\ref{fig:omegaVSconcurrence}  and~\ref{fig:omegaVSconcurrencet0} in which $\Theta_{\rm GW}/A \lambda^2$ is plotted as a function of the gravitational wave frequency $\omega \sigma$ for different detector energies $\Omega \sigma$ for both spacelike and timelike separated detectors. From Fig.~\ref{fig:omegaVSconcurrence}, we see that for both spacelike and timelike separated detectors $\Theta_{ \rm  GW}$ is a negative quantity, supporting the conclusion that gravitational waves degrade field entanglement for $t_0 = 0$, as described in the previous paragraph.  Moreover, Fig.~\ref{fig:omegaVSconcurrence} reveals a strong resonance effect when the frequency of the gravitational wave is approximately equal to the energy gap of the detector, $\omega \approx 2 \Omega$, around which the harvested entanglement is maximally degraded. This resonance is due to the dependence of $\Theta_{\rm GW}$ on the Gaussian profile centered at $\omega = 2 \Omega$ that appears in Eq.~\eqref{functionf}. Away from this resonance, $\Theta_{\rm GW}$ approaches zero asymptotically, which implies that the gravitational wave does not influence the harvested entanglement significantly when $\abs{\omega - \Omega } \gg 1 / \sigma $.  Note that if the atom had begun in its excited state, $\Omega \to -\Omega$, then $\Theta_{\rm GW}$ would be identical, which implies that for $t_0=0$ the harvested entanglement would be degraded by the same amount.

In contrast, Fig.~\ref{fig:omegaVSconcurrencet0} depicts $\Theta_{\rm GW}$ when $t_0 \neq 0$, revealing oscillatory behaviour of the concurrence as a function of $\omega$ around the resonance condition $\omega \approx 2\Omega$. The frequency of these oscillations is $t_0$, which can be seen by expanding the numerator in Eq.~\eqref{GWcontributiontoconcrrrence} and noting that it is a sum of terms that oscillate with this frequency. It is thus seen that $\Theta_{\rm GW}$ can be positive or negative, indicating that a gravitational wave can either amplify or degrade the harvested entanglement depending on $\omega \sigma$ and $t_0/\sigma$. Again, when $\omega$ moves away from $\omega \approx 2\Omega$, $\Theta_{\rm GW}$  approaches zero asymptotically.

The effect a gravitational wave has on the total correlations harvested by a pair of detectors is discussed in Appendix~\ref{corr}, revealing that harvested correlations are affected in a similar fashion as harvested entanglement.

\section{Conclusion and outlook}
\label{conclusion}

We examined the effect that a gravitational wave has on Unruh-DeWitt detectors. To do so, the Wightman function for a massless scalar field living in a gravitational wave background was derived and used to compute the final states of one and two detectors locally coupled to the field for a finite period of time.

It was shown that the transition probability of an inertial detector is unaffected by a gravitational wave, in agreement with Gibbon's  observation that a gravitational wave does not excite particles from the vacuum~\cite{gibbonsQuantizedFieldsPropagating1975}. In contrast, the entanglement structure of the vacuum was shown to be modified by the presence of a gravitational wave as witnessed by the entanglement harvesting protocol. When the detectors are tuned to the frequency of the gravitational wave, it was shown that depending on when the detectors interact with the field relative to where the gravitational wave is in its cycle, the harvested entanglement can be either amplified or degraded relative to an equivalent pair of detectors in Minkowski space.

The relative size of the gravitational wave contribution to the entanglement harvested, $ |\Theta_{\rm GW} / \Theta_{\mathcal{M}}|$, is proportional to the amplitude of the gravitational wave. Since our analysis was carried out in the linearized gravity regime, it would be interesting to extend the analysis to the strong gravity regime where similar resonance effects would presumably exist, which may generate a more easily detectable gravitational wave signal. Moreover, different detector configurations could potentially yield further amplification of harvested entanglement. Furthermore, in the strong gravity regime it would be interesting to examine the consequences of gravitational-wave memory effect~\cite{Christodoulou1991nonlinear, Wisemanmem1991} on vacuum entanglement, revealing potential differences in the way in which classical and quantum systems are affected. One might also imagine extending this analysis to investigate gravitational-wave induced decoherence; since one cannot shield from gravity, such a decoherence mechanism might be expected to affect all systems.

\begin{acknowledgments}
We thank Robert B. Mann and Eduardo Mart\'{i}n-Mart\'{i}nez for useful comments. This work was supported by the Natural Sciences and Engineering Research Council of Canada and the Dartmouth Society of Fellows.  \end{acknowledgments}

\bibliography{DetectorsAndGW}

\onecolumngrid

\appendix

\section{Derivation of gravitational wave spacetime Wightman function}
\label{Derivation of Wightman function}

Consider a massless scalar field $\phi(\mathsf{x}) =\phi(u,v,x,y)$ in a gravitational wave background satisfying the Klein-Gordon equation $\Box \phi(x) = 0$  in Eq.~\eqref{KGequation}. The Klein-Gordon equation is separable in the coordinates $(u,v,x,y)$ and an arbitrary solution can be expanded in the complete set of mode functions
\begin{align}
u_{\vec{k}}(u,v,x,y) = \frac{\gamma^{-1}(u)}{\sqrt{2k_{-}}(2\pi)^{3/2}} \exp\left [ik_x x + ik_y y -ik_{-}v -\frac{i}{4k_{-}} \int_0^u du \, g^{ab}k_a k_b  \right] \nn.
\end{align}
where $\gamma(u) \ce (1 - A^2 \cos \omega u)^{1/4}$ and the integral evaluates to
\begin{align}
 \int_0^u du \, g^{ab}k_a k_b 
&= \int_0^u du \, \left[ k_x^2(1- A \cos \omega u) + k_y^2(1 + A \cos \omega u) \right] \nn \\
&= \left( k_x^2 + k_y^2 \right) u - \left( k_x^2 - k_y^2 \right)\frac{A}{\omega} \sin \omega u. \nn
\end{align}
These mode functions are normalized and orthogonal to one another with respect to the usual Klein-Gordon inner product~\cite{garrigaScatteringQuantumParticles1991, BirrellDavies}. The Wightman function $W(x,x') \ce \braket{0|\phi(x) \phi(x') | 0}$ can be expressed in terms of these mode as
\begin{align}
W(\mathsf{x},\mathsf{x}') 
&= \int d \vec{k} \, u_{\vec{k}}(u,v,x,y),u_{\vec{k}'}^*(u',v',x',y') \nn \\
&=\int d \vec{k} \frac{\gamma^{-1}(u)\gamma^{-1}(u')}{(2\pi)^3 2 k_{-}  } e^{i k_x \Delta x + ik_y \Delta y-i k_{-} \Delta v - \frac{i}{4 k_-}\left( k_x^2 + k_y^2 \right) \Delta u + \frac{i}{4k_-} \left( k_x^2 - k_y^2 \right)\frac{A}{\omega} \left( \sin \omega u  - \sin \omega u' \right)   } \nn \\
&=\int d \vec{k} \frac{\gamma^{-1}(u)\gamma^{-1}(u')}{(2\pi)^3 2 k_{-}  } e^{i k_x \Delta x + ik_y \Delta y-i k_{-} \Delta v - \frac{ik_x^2}{4 k_-}\left[ \Delta u  - \frac{2A}{\omega} \sin \left(\omega \frac{\Delta u}2 \right) \cos \left( \omega \frac{u+u'}{2} \right) \right]   - \frac{ik_y^2}{4 k_-}\left[ \Delta u  + \frac{2A}{\omega} \sin \left(\frac{\Delta u}2 \right) \cos \left( \frac{u+u'}{2} \right) \right]   } . \nn
\end{align}
Expanding to leading order in $A$ yields
\begin{align}
W(\mathsf{x},\mathsf{x}') 
&=\int \frac{d \mathsf{k}}{(2\pi)^3 2 k_{-}  } e^{i k_x \Delta x + ik_y \Delta y-i k_{-} \Delta v - \frac{i}{4 k_-} \left(k_x^2+ k_y^2 \right) \Delta u  }\left[ 1 +\frac{i A}{2 \omega} \frac{k_x^2- k_y^2 }{k_-} \sin \left(\tfrac{\omega}{2} \Delta u \right)  \cos \left( \tfrac{\omega}{2} [u+u'] \right) \right]. \nn
\end{align}
The first term yields the Minkowski space Wightman function 
\begin{align}
W_{\mathcal{M}} (\mathsf{x},\mathsf{x}') &= \frac{1}{4\pi i \Delta u  }    \delta\left(\frac{\sigma_{\mathcal{M}}(\mathsf{x},\mathsf{x}')}{\Delta u}  \right)  
+ \PV \frac{1}{4\pi^2  \sigma_{\mathcal{M}}(\mathsf{x},\mathsf{x}')} , \nn
\end{align}
and the second term evaluates to
\begin{align*}
    W_{\rm GW} (\mathsf{x},\mathsf{x}') &= \frac{iA}{(2\pi)^3 4 \omega   } \sin \left(\tfrac{\omega}{2} \Delta u \right)  \cos \left( \tfrac{\omega}{2} [u+u'] \right) \int dk_x dk_y dk_- \, e^{i k_x \Delta x + ik_y \Delta y-i k_{-} \Delta v - \frac{i}{4 k_-} \left(k_x^2+ k_y^2 \right) \Delta u  }  \frac{k_x^2- k_y^2 }{k_-^2} \\
    &= \frac{A}{2 \omega \pi^2 } \frac{\Delta x^2 - \Delta y^2}{\Delta u^3} \sin{\frac{\omega \Delta u}{2}} \cos{\frac{\omega (u + u')}{2}} \int dk_{-} k_{-} e^{ik_{-}\left( -\Delta v + \frac{\Delta x^2}{\Delta u} + \frac{\Delta y^2}{\Delta u} \right)}. \nn
\end{align*}
To evaluate the last integral, consider a function $f = f(x)$ and the following integral
\begin{align}
    \int_{0}^{\infty} dx \, x e^{i f x}= -i \frac{d}{df} \int_{0}^{\infty} dx \,  e^{i f x}= -i \frac{d}{df}\left[ \pi \delta(f) + \PV \frac{i}{f}\right]= -\left[ i \delta'(f) + \PV \frac{1}{f^2}\right]. \nn
\end{align}
Then, the gravitational wave Wightman function becomes
\begin{align}
   W_{\rm GW}(\mathsf{x},\mathsf{x}') =  -\frac{ A}{ 4\pi^2} \frac{\sin \left(\tfrac{\omega}{2} \Delta u \right) }{\tfrac{\omega}{2}  \Delta u } \cos \left( \tfrac{\omega}{2} [u+u'] \right) \frac{\Delta x^2 - \Delta y^2}{\Delta u^2}   \left[ i \pi \delta' \left( \frac{\sigma_{\mathcal{M}}(\mathsf{x},\mathsf{x}')}{\Delta u}\right)
+ \PV \frac{ \Delta u^2}{\sigma_{\mathcal{M}}^2(\mathsf{x},\mathsf{x}')} \right]. \nn
\end{align}

\section{Computing $P$, $X$ and $C$}
\label{Derivation of P, X and C}
\begin{center}
    \emph{Derivation of $P$}
\end{center}
Recall from Eq.~\eqref{tprob} that the probability $P$ for a detector to transition from its ground state to its excited state to leading order in the interaction strength is 
\begin{align}
P &= \lambda^2 \int dt  d t' \, \chi(t) \chi(t') e^{-i \Omega \left(t-t'\right)} W\!\left(\mathsf{x}_D(t) , \mathsf{x}_D(t')\right), \nn
\end{align}
Substituting in the explicit form of the switching functions, it follows that
\begin{align}
P = \lambda^2 \int dt \int dt' e^{-\frac{(t-t_0)^2 + (t'-t_0)^2}{2\sigma^2}}e^{-i \Omega (t-t')} W(\mathsf{x}(t),\mathsf{x}(t')). \nn
\end{align}
Consider the trajectory of a single detector in Eq.~\eqref{SingleTrajectory}; since $\Delta x  = \Delta y = 0$, we immediately see that the gravitational wave contribution to the Wightman function in Eq.~\eqref{WGW} vanishes. Thus, the transition probability of a single detector is unaffected by the presence of a gravitational wave. To evaluate the transition probability, consider the change of variable $a \ce \Delta t = t-t'$ and $b \ce t+t'$, yielding 
\begin{align}
P &= \frac{1}{2} \lambda^2 \int da \int db \,  e^{-\frac{a^2 +(b-2 t_0)^2}{4 \sigma^2}} e^{-i\Omega a} \left [ \frac{1}{4\pi i a  }    \delta\left(-a \right)  
+ \PV \frac{1}{4\pi^2(-a^2)} \right] \nn \\
& = \lambda^2 \sigma \sqrt{\pi} \int da \,  e^{\frac{-a^2}{4 \sigma^2}} e^{-i\Omega a} \left [ \frac{1}{4\pi i a  }    \delta\left(a \right)  
+ \PV \frac{1}{4\pi^2(-a^2)} \right] \nn \\
& = \lambda^2 \sigma \sqrt{\pi} \left [\frac{-\Omega}{4\pi} + \frac{1}{4\pi \sqrt{\pi} \sigma} e^{-\sigma^2 \Omega^2} +\frac{\Omega \erf(\sigma \Omega)}{4\pi} \right] \nn \\
& = \frac{\lambda^2}{4 \pi} \left[ e^{-\sigma^2 \Omega^2} - \sqrt{\pi} \sigma \Omega \erfc(\sigma \Omega)\right]. \nn
\end{align}
The second last equality follows from the distribution identities: $\frac{\delta(x)}{x} = -\delta'(x)$ and 
\begin{align}
\PV \int_{\infty}^{\infty} dx \, \frac{f(x)}{x^2} = \int_{0}^{\infty} dx \, \frac{f(x)+f(-x)-2 f(0)}{x^2}, \nn
\end{align}
where it is assumed $f(x)$ reaches 0 as $x \rightarrow \pm \infty$. \\

\begin{center}
    \emph{Derivation of $X_{\rm \mathcal{M}}$}
\end{center}
The matrix element is given by
\begin{align}
X_{\rm \mathcal{M}} = -\lambda^2 \int_{-\infty}^\infty dt \int_{-\infty}^t dt' \, e^{-\frac{(t-t_0)^2 + (t-t_0)^2}{2\sigma^2}} e^{-i\Omega (t + t')}\left [ W_{\rm \mathcal{M}}(\mathsf{x}_A(t'), \mathsf{x}_B(t)) + W_{\rm \mathcal{M}}(\mathsf{x}_B(t'),\mathsf{x}_A(t))        \right]. \nn
\end{align}
The Wightman function for Minkowski space for our trajectories becomes
\begin{align}
 W_{\rm \mathcal{M}}(\mathsf{x}_A(t'), \mathsf{x}_B(t))=   -\frac{1}{4\pi i \Delta t  }    \delta\left( \Delta t - \frac{D^{2}}{\Delta t} \right)  
+ \PV \frac{1}{4\pi^2 (-\Delta t^{2} +D^{2})}. \nn
\end{align}
By changing variables to $a=\Delta t,b=t+t'$, we find the matrix element $X$ in Minkowski space
\begin{align}
 X_{\rm \mathcal{M}} &= -2\lambda^2 \int_{-\infty}^\infty dt \int_{-\infty}^t dt' \, e^{-\frac{(t -t_{0})^2 + (t'-t_{0})^2}{2\sigma^2}} e^{-i\Omega (t + t')}\left[-\frac{1}{4\pi i \Delta t  }    \delta\left( \Delta t - \frac{D^{2}}{\Delta t} \right)  
+ \PV \frac{1}{4\pi^2 (-\Delta t^{2} +D^{2})}\right] \nn \\
&= -\lambda^2 e^{-2i\Omega t_{0}} \int_{-\infty}^\infty db e^{-\frac{(b-2t_{0}) ^2}{4\sigma^2}-i\Omega b} \int_{0}^{\infty} da \, e^{-\frac{a ^2}{4\sigma^2}} \left[\frac{1}{4\pi i a  }    \delta\left( a - \frac{D^{2}}{a} \right)
+ \PV \frac{1}{4\pi^2 (a^{2} -D^{2})}\right] \nn \\
&=2 \sigma \sqrt{\pi}\lambda^2 e^{-\Omega^{2}\sigma^{2}-2i\Omega t_{0}} \int_{0}^{\infty} da \, e^{-\frac{a ^2}{4\sigma^2}} \left[\frac{1}{4\pi i a  }    \delta\left( a - \frac{D^{2}}{a} \right)
+ \PV \frac{1}{4\pi^2 (a^{2} -D^{2})}\right] \nn \\
&= i \frac{\lambda^2 \sigma}{4 D \sqrt{\pi}} e^{-\sigma^{2}\Omega^2 -2i\Omega t_{0}- \frac{D^{2}}{ 4\sigma^{2}}} \left[ \erf \left( i \frac{D}{2\sigma}\right) -1 \right]. \nn
\end{align}
where the principal value integration was evaluated using methods similar to those in~\cite{smithDetectorsReferenceFrames2019}.

\begin{center}
    \emph{Derivation of $X_{\rm GW}$}
\end{center}

The matrix element $X$ is given by~\cite{reznikViolatingBellInequalities2005,pozas-kerstjensEntanglementHarvestingElectromagnetic2016,smithDetectorsReferenceFrames2019,martin-martinezSpacetimeStructureVacuum2016}
\begin{align}
X_{\rm GW} = -\lambda^2 \int_{-\infty}^\infty dt \int_{-\infty}^t dt' \, e^{-\frac{t ^2 + t'^2}{2\sigma^2}} e^{-i\Omega (t + t')}\left [ W_{\rm GW}(\mathsf{x}_A(t'), \mathsf{x}_B(t)) + W_{\rm GW}(\mathsf{x}_B(t'),\mathsf{x}_A(t))        \right]. \nn
\end{align}
From Eq.~\eqref{twodetectortraj}, it is seen that
$\sigma_{\mathcal{M}}(\mathsf{x}_A(t'), \mathsf{x}_B(t)) = \sigma_{\mathcal{M}}(\mathsf{x}_B(t'), \mathsf{x}_A(t) )= - \Delta t^2 + D^2$. It follows
\begin{align}
 W_{\rm GW}(\mathsf{x}_A(t'), \mathsf{x}_B(t)) &=W_{\rm GW}(\mathsf{x}_B(t'), \mathsf{x}_A(t)) \nn \\
 &=  -\frac{ A}{ 4\pi^2} \frac{\sin \left(\tfrac{\omega}{2} \Delta t \right) }{\tfrac{\omega}{2} \Delta t } \cos \left( \tfrac{\omega}{2} [t+t'] \right) \frac{D^2}{\Delta t^2}   \left[ i\pi \delta' \left(  \Delta t-\frac{D^2}{\Delta t} \right)
+ \PV\left( \frac{\Delta t}{D^2 - \Delta t^2} \right)^{2} \right]. 
\label{EvaluatedWGW}
\end{align}
which we note is invariant under $t \leftrightarrow t'$. It follows that $X$ may be expressed as
\begin{align}
X_{\rm GW} &= A \frac{ \lambda^2 D^2}{ 2\pi^2}\int_{-\infty}^\infty dt \int_{-\infty}^t dt' \,   e^{-i\Omega (t' + t)}  e^{-\frac{(t'+ t -2t_{0})^2}{4\sigma^2}}  \cos \left( \tfrac{\omega}{2} [t+t'] \right)\nn \\
&\hspace{2.25in} \times \frac{e^{-\frac{\Delta t^2}{4\sigma^2}} }{\Delta t^2}  \frac{\sin \left(\tfrac{\omega}{2} \Delta t \right) }{\tfrac{\omega}{2} \Delta t }      \left[ i\pi \delta' \left(  \Delta t- \frac{D^2}{\Delta t} \right)
+ \PV\left( \frac{\Delta t}{D^2 - \Delta t^2} \right)^{2} \right].\nn
\end{align}
Changing integration variables to $a \ce \Delta t$ and $b\ce t'+ t $, yields
\begin{align}
X_{\rm GW} &=  \frac{ A \lambda^2 D^2}{ 4\pi^2}  \int_{-\infty}^\infty db \, e^{-i\Omega b } e^{-\frac{(b-2t_{0})^2}{4\sigma^2}}\cos \left( \tfrac{\omega}{2} b  \right) \int_{0}^{\infty}  da  \,    \frac{e^{-\frac{a ^2}{4\sigma^2}} }{a^2} \frac{\sin \left(\tfrac{\omega}{2} a \right) }{\tfrac{\omega}{2} a}   \left[ i\pi \delta' \left(   a- \frac{D^2}{a} \right)
+ \PV\left( \frac{a}{D^2 - a^2} \right)^{2} \right] \nn \\
& =  \frac{ A \lambda^2 D^2}{ 2\pi^2} \sqrt{\pi }\sigma e^{- \left( \frac{ \sigma ^{2}\omega^2}{4} + \sigma^{2}\Omega^2\right)} e^{-2i t_0 \Omega} \cosh \left( \omega \Omega \sigma^2 -it_0 \omega \right) \nn \\
&\quad \times \int_{0}^{\infty}  da  \,    \frac{e^{-\frac{a ^2}{4\sigma^2}} }{a^2} \frac{\sin \left(\tfrac{\omega}{2} a \right) }{\tfrac{\omega}{2} a}   \left[ i\pi \delta' \left(   a- \frac{D^2}{a} \right)
+ \PV\left( \frac{a}{D^2 - a^2} \right)^{2}  \right] \nn \\
& =  \frac{ A \lambda^2 }{ 2 D^2 \pi^2 } \sqrt{\pi }\sigma e^{- \left( \frac{ \sigma ^{2}\omega^2}{4} + \sigma^{2}\Omega^2 + 2i t_0 \Omega\right)}  \cosh \left( \omega \Omega \sigma^2 -it_0 \omega \right)  \left(I_1 + I_2 \right),
\label{XGW1}
\end{align}
where the last equality defines the $I_1$ and $I_2$ that remain to be evaluated. To evaluate the first integral in Eq.~\eqref{XGW1}, note that 
\begin{align}
 \frac{d}{da} \delta \left(   a- \frac{D^2}{a} \right) =  \delta' \left(   a- \frac{D^2}{a} \right) \left(\frac{D^2}{a^2}+1\right) \implies \delta' \left(   a- \frac{D^2}{a} \right)  =  \left[\frac{d}{da} \delta \left(   a- \frac{D^2}{a} \right) \right] \left(\frac{D^2}{a^2}+1\right)^{-1}.\nn
\end{align}
Then,
\begin{align}
 I_1 &\ce i D^4 \pi  \int_{0}^{\infty}  da  \,    \frac{e^{-\frac{a ^2}{4\sigma^2}} }{a^2} \frac{\sin \left(\tfrac{\omega}{2} a \right) }{\tfrac{\omega}{2} a}  \delta' \left(   a- \frac{D^2}{a} \right) \nn \\
 &= i D^4\pi \int_{0}^{\infty}  da  \,   \left[\frac{d}{da} \delta \left(   a- \frac{D^2}{a} \right) \right] \left(\frac{D^2}{a^2}+1\right)^{-1}  \frac{e^{-\frac{a ^2}{4\sigma^2}} }{a^2} \frac{\sin \left(\tfrac{\omega}{2} a \right) }{\tfrac{\omega}{2} a}  \nn \\
  &= -i D^4\pi \int_{0}^{\infty}  da  \,   \delta\left( \frac{D^2}{a}- a \right) \frac{d}{da}  \left[ \left(\frac{D^2}{a^2}+1\right)^{-1}  \frac{e^{-\frac{a ^2}{4\sigma^2}} }{a^2} \frac{\sin \left(\tfrac{\omega}{2} a \right) }{\tfrac{\omega}{2} a} \right] \nn \\
   &= -i D^4\pi \int_{0}^{\infty}  da  \,   \frac{\delta\left( D - a \right)}{2} \frac{d}{da}\left[ \left(\frac{D^2}{a^2}+1\right)^{-1}  \frac{e^{-\frac{a ^2}{4\sigma^2}} }{a^2} \frac{\sin \left(\tfrac{\omega}{2} a \right) }{\tfrac{\omega}{2} a} \right]\nn \\
   &= i \frac{ \pi e^{-\frac{ D^2 }{ 4\sigma^{2}}}}{ \omega} \left[\left(\frac{D^{2}}{4\sigma^{2}}+1 \right)\sin\left( \tfrac{ \omega}{2} D\right)- \frac{D  \omega}{4} \cos \left( \tfrac{ \omega}{2} D\right)     \right] . \nn
\end{align}
Next, evaluating the second integral in Eq.~\ref{XGW1} yields
\begin{align}
    I_2 &\ce D^4
\PV \int_{0}^{\infty}  da  \,    e^{-\frac{a ^2}{4\sigma^2}}  \frac{\sin \left(\tfrac{\omega}{2} a \right) }{\tfrac{\omega}{2} a}  \left( D^2 - a^2 \right)^{-2} \nn \\
&= D^4 \PV \int_{-\infty}^{\infty}  da  \,    e^{-\frac{a ^2}{4\sigma^2}}  \frac{\sin \left(\tfrac{\omega}{2} a \right) }{\omega a}  \left( D^2 - a^2 \right)^{-2} \nn \\
&=  \frac{D^4}{\omega} \PV \int_{-\infty}^{\infty}  da \, e^{-\frac{a^2}{4\sigma^2}  } \sin \left(\tfrac{\omega}{2} a \right) \int_{-\infty}^{\infty} d\bar{a}\, \delta (\bar{a} - a)  \frac{1}{\bar{a} (\bar{a}^2 - D^2)^{2}} \nn \\
&=  \frac{D^4}{\omega} \PV \int_{-\infty}^{\infty}  da \, e^{-\frac{a^2}{4\sigma^2}  } \sin \left(\tfrac{\omega}{2} a \right) \int_{-\infty}^{\infty} d\bar{a}\, \left( \frac{1}{2 \pi} \int_{-\infty}^{\infty} ds \, e^{i (\bar{a}-a) s}  \right)\frac{1}{\bar{a} (\bar{a}^2 - D^2)^{2}} \nn \\
&=   \frac{D^4}{2 \pi \omega}  \PV \int_{-\infty}^{\infty} ds  \left[ \int_{-\infty}^{\infty}  da \, e^{-i a s} e^{-\frac{a^2}{4\sigma^2}  } \sin \left(\tfrac{\omega}{2} a \right) \right] \left[ \int_{-\infty}^{\infty} d\bar{a}\,  e^{i \bar{a} s}  \frac{1}{\bar{a} (\bar{a}^2 - D^2)^{2}}\right] \nn \\
&=  \frac{1}{\omega} \int_{-\infty}^{\infty} ds  \left[  -2 \sqrt{\pi}i \sigma e^{-\sigma^{2}\left( s^2 + \frac{\omega^{2}}{4}\right)} \sinh{(\sigma^{2} \omega s)}   \right] \left[ \frac{i \sgn\left(s\right) }{4} \left( 2 -2 \cos [D s] - D s \sin [D s] \right)\right] \nn \\
&=  \frac{\sqrt{\pi} \sigma}{2 \omega} e^{-\left( \frac{ \sigma\omega }{2} \right)^2 }  \int_{-\infty}^{\infty} ds  \, \sgn\left(s\right)   e^{-\sigma^{2} s^2  } \sinh \left (\sigma^{2} \omega s \right)    \left[ 2 -2 \cos \left( D s \right) - D s \sin \left(D s\right) \right] \nn \\
  &= \frac{\pi }{ \omega} \left(   \erf \left(\frac{\sigma  \omega }{2}\right) - 
 e^{-\frac{D^2}{4\sigma^2}}  \Re \left[e^{i \frac{\omega}{2} D } \left(1 + \frac{D^2}{4 \sigma^2}-i \frac{D  \omega}{4}  \right) \erf \left( \frac{\omega}{2} \sigma +  \frac{ i D }{2 \sigma } \right) \right] \right). \nn
\end{align}
\begin{center}
    \emph{Derivation of $C_{\rm \mathcal{M}}$}
\end{center}
The expression for $C_{\rm \mathcal{M}}$ is the following
\begin{align}
C_{\rm \mathcal{M}} &= \lambda^2 \int_{-\infty}^{\infty} dt \int_{-\infty}^{\infty} dt' \,  e^{-\frac{(t-t_{0}) ^2 + (t'-t_{0})^2}{2\sigma^2}} e^{i\Omega (t - t')}W_{\rm \mathcal{M}}(\mathsf{x}_{A}(t'),\mathsf{x}_{B}(t)). \nn
\end{align}
By plugging in the Wightman function in Minkowski space for the trajectories of the detectors and then changing variables to $a=\Delta t, b=t+t'$, we obtain 
\begin{align}
    C_{\rm \mathcal{M}} &= \frac{\lambda^{2}}{2} \int_{-\infty}^{\infty}db e^{-\frac{(b-2t_{0})^2}{4\sigma^{2}}} \int_{-\infty}^{\infty} da e^{-\frac{a^2}{4\sigma^{2}}+i \Omega a}\left[-\frac{1}{4\pi i a  }    \delta\left( a - \frac{D^{2}}{a} \right)  
+ \PV \frac{1}{4\pi^2 (-a^{2} +D^{2})}\right] \nn \\
&=-\sigma \sqrt{\pi}\lambda^{2} \int_{-\infty}^{\infty} da e^{-\frac{a^2}{4\sigma^{2}}+i \Omega a}\left[\frac{1}{4\pi i a  }    \delta\left( a - \frac{D^{2}}{a} \right)  
+ \PV \frac{1}{4\pi^2 (a^{2} -D^{2})}\right] \nn \\
&=\sigma \sqrt{\pi}\lambda^{2}e^{-\frac{D^2}{4\sigma^{2}}} \left[ \frac{\sin(\Omega D)}{4 D \pi}+ \frac{1}{4D \pi} \Re \left(ie^{i D \Omega} \erf{\left[i\frac{D}{2\sigma} +\sigma\Omega\right]}\right)\right] \nn \\
&= \frac{\sigma \lambda^2}{4 D\sqrt{\pi}} e^{-\frac{D^{2}}{ 4\sigma^{2}} } \left(\Im \left[ e^{i D \Omega} \erf \left( i \frac{D}{2\sigma} +  \sigma \Omega \right)\right]   - \sin  \Omega D \right), \nn
\end{align}
where the principal value integration was evaluated using methods similar to those in~\cite{smithDetectorsReferenceFrames2019}.

\begin{center}
    \emph{Derivation of $C_{\rm GW}$}
\end{center}

The expression for $C_{\rm GW}$ is given by~\cite{reznikViolatingBellInequalities2005,pozas-kerstjensEntanglementHarvestingElectromagnetic2016,smithDetectorsReferenceFrames2019,martin-martinezSpacetimeStructureVacuum2016}
\begin{align}
C_{\rm GW} &= \lambda^2 \int_{-\infty}^{\infty} dt \int_{-\infty}^{\infty} dt' \,  e^{-\frac{(t-t_{0}) ^2 + (t'-t_{0})^2}{2\sigma^2}} e^{i\Omega (t - t')}W_{\rm GW}(\mathsf{x}_{A}(t'),\mathsf{x}_{B}(t)).
\label{C1simp}
\end{align}
Using Eq.~\eqref{EvaluatedWGW} and changing integration variables to $a\ce \Delta t$ and $b \ce t+t'$, Eq.~\eqref{C1simp} becomes
 \begin{align}
C_{\rm GW} 
&= -\frac{  \lambda^2 A D^2 }{ 4\pi^2}   \int_{-\infty}^{\infty} dt \int_{-\infty}^{\infty} dt' \,  e^{-\frac{\left(t+t' -2t_{0}\right)^{2}}{4\sigma^2}}  \cos \left( \tfrac{\omega}{2} [t+t'] \right)  \nn \\
&\hspace{2.25in} \times e^{i\Omega  \Delta t } \frac{e^{-\frac{ \Delta t ^2}{4\sigma^2}}  }{\Delta t^2}    \frac{\sin \left(\tfrac{\omega}{2} \Delta t \right) }{\tfrac{\omega}{2} \Delta t }    \left[ i\pi \delta' \left(  \Delta t-\frac{D^2}{\Delta t} \right)
+ \PV\left( \frac{\Delta t}{D^2 - \Delta t^2} \right)^{2} \right] \nn \\
&= -\frac{  \lambda^2 A D^2 }{ 8\pi^2}   \int_{-\infty}^{\infty} db  \,  e^{-\frac{(b-2t_{0})^2}{4\sigma^2}}  \cos \left( \tfrac{\omega}{2} b \right)     \int_{-\infty}^{\infty} da \, e^{i\Omega  a } \frac{e^{-\frac{ a^2}{4\sigma^2}}  }{a^2}    \frac{\sin \left(\tfrac{\omega}{2} a \right) }{\tfrac{\omega}{2} a }    \left[ i\pi \delta' \left(  a-\frac{D^2}{a} \right)
+ \PV\left( \frac{a}{D^2 - a^2} \right)^{2} \right] \nn \\
&= -\frac{  \lambda^2 A D^2 }{ 8\pi^2}  \left[2 \sqrt{\pi} \sigma e^{- \left( \frac{\omega}{2} \sigma \right)^2} \cos \left(\omega t_{0}\right)\right]     \int_{-\infty}^{\infty} da \, e^{i\Omega  a } \frac{e^{-\frac{ a^2}{4\sigma^2}}  }{a^2}    \frac{\sin \left(\tfrac{\omega}{2} a \right) }{\tfrac{\omega}{2} a }    \left[ i\pi \delta' \left(  a-\frac{D^2}{a} \right)
+ \PV\left( \frac{a}{D^2 - a^2} \right)^{2} \right] \nn \\
&= -\frac{  \lambda^2 A \sigma }{ 4 D^{2} \pi^{3/2}}    e^{- \left( \frac{\omega}{2} \sigma \right)^2}    \cos \left(\omega t_{0}\right)  \left( I_3 + I_4 \right), \nn
\end{align}
where
\begin{align}
I_3 &\ce   i D^4\pi  \int_{-\infty}^{\infty} da \, e^{i\Omega  a } \frac{e^{-\frac{ a^2}{4\sigma^2}}  }{a^2}    \frac{\sin \left(\tfrac{\omega}{2} a \right) }{\tfrac{\omega}{2} a }    \delta' \left(  a-\frac{D^2}{a} \right) 
\nn \\
&=     iD^4\pi  \int_{-\infty}^{\infty} da \, e^{i\Omega  a } \frac{e^{-\frac{ a^2}{4\sigma^2}}  }{a^2}    \frac{\sin \left(\tfrac{\omega}{2} a \right) }{\tfrac{\omega}{2} a }   \left[\frac{d}{da} \delta \left(  a-\frac{D^2}{a} \right) \right] \left(\frac{D^2}{a^2}+1\right)^{-1} \nn \\
&=    -iD^4\pi  \int_{-\infty}^{\infty} da \,  \delta \left(  a-\frac{D^2}{a} \right)  \frac{d}{da} \left[  e^{i\Omega  a } \frac{e^{-\frac{ a^2}{4\sigma^2}}  }{a^2}    \frac{\sin \left(\tfrac{\omega}{2} a \right) }{\tfrac{\omega}{2} a }    \left(\frac{D^2}{a^2}+1\right)^{-1} \right] \nn \\
&=    -i  D^4\pi \int_{-\infty}^{\infty} da \,    \frac{\delta\left( D + a \right) +   \delta\left( D-a \right)}{2} \frac{d}{da} \left[  e^{i\Omega  a } \frac{e^{-\frac{ a^2}{4\sigma^2}}  }{a^2}    \frac{\sin \left(\tfrac{\omega}{2} a \right) }{\tfrac{\omega}{2} a }    \left(\frac{D^2}{a^2}+1\right)^{-1} \right] \nn \\
&=  \frac{ \pi e^{- \frac{D^2}{4 \sigma^2}}}{2  \omega }  \left[   D  \omega  \sin \left( \Omega D \right) \cos \left( \tfrac{\omega}{2} D  \right) +  2D  \Omega \cos \left( \Omega D \right) \sin\left( \tfrac{\omega}{2} D \right)   -\left( \frac{D^{2}}{\sigma^{2}} + 4  \right)  \sin \left( \Omega D \right)\sin\left( \tfrac{\omega}{2} D \right)  \right] \nn \\
&= \frac{ \pi e^{- \frac{D^2}{4 \sigma^2}}}{4  \omega } \Bigg[\left(D \omega +2 D \Omega\right) \sin \left(D\left[\frac{\omega}{2}+\Omega\right]\right) +  \left(D \omega -2 D \Omega\right) \sin \left(D\left[\frac{\omega}{2}-\Omega\right]\right) \nn \\ 
    & + \left(\frac{D^{2}}{\sigma^{2}}+4\right)\left( \cos \left(D \left[ \frac{\omega}{2} + \Omega\right]\right) - \cos \left(D \left[\frac{\omega}{2} - \Omega\right]\right)\right)\Bigg] \nn
\end{align}
and
\begin{align}
I_4 &\ce  D^4\PV \int_{-\infty}^{\infty} da \, e^{i\Omega  a }      \frac{\sin \left(\tfrac{\omega}{2} a \right) }{\tfrac{\omega}{2} a }    \frac{e^{-\frac{ a^2}{4\sigma^2}}  }{\left( a^2 - D^2\right)^{2}}   \nn \\
&=  \frac{2D^4}{\omega} \PV \int_{-\infty}^{\infty} da \, e^{i\Omega  a }      \sin \left(\tfrac{\omega}{2} a\right)   e^{-\frac{ a^2}{4\sigma^2}}     \int_{-\infty}^{\infty} d\bar{a} \, \delta(\bar{a}-a)  \frac{1 }{ \bar{a} \left(  \bar{a}^2 - D^2\right)^{2}} \nn \\ 
&=  \frac{2D^4}{\omega} \PV \int_{-\infty}^{\infty} da \, e^{i\Omega  a }      \sin \left(\tfrac{\omega}{2} a\right)   e^{-\frac{ a^2}{4\sigma^2}}     \int_{-\infty}^{\infty} d\bar{a}  \left( \frac{1}{2 \pi} \int_{-\infty}^{\infty} ds \, e^{i (\bar{a}-a) s}  \right)    \frac{1 }{ \bar{a} \left(  \bar{a}^2 - D^2\right)^{2}}    \nn \\
&=  \frac{D^4}{ \pi \omega} \PV \int_{-\infty}^{\infty} ds   \int_{-\infty}^{\infty} da \,  e^{i(\Omega -s) a }      \sin \left(\tfrac{\omega}{2} a\right)   e^{-\frac{ a^2}{4\sigma^2}}     \int_{-\infty}^{\infty} d\bar{a} \,   \frac{ e^{i \bar{a} s } }{ \bar{a} \left(  \bar{a}^2 - D^2\right)^{2}}    \nn \\
 &=  \frac{\sqrt{\pi} \sigma}{  \omega } e^{- \frac{\sigma^2 \omega^2}{4} }  \int_{-\infty}^{\infty} ds   \left[  e^{- \sigma^2 (\Omega-s)^2 } \sinh \left( \sigma^2  \omega (\Omega-s)  \right) \right]  
 \left[ \sgn \left(s \right) \left(D s  \sin \left(D s \right)+2 \cos \left(D s \right)-2 \right) \right]   \nn \\
    &=\frac{\pi}{ \omega}\left( \erf \left[\sigma \left( \frac{ \omega}{2}-\Omega \right) \right]+\erf\left[\sigma\left( \frac{ \omega}{2}+\Omega \right)\right]
    -  e^{-\frac{D^{2}}{4\sigma^{2}}}\Re\left[ Q_{+}R_{+} +  Q_{-}R_{-} \right] \right), 
\end{align}
where we have defined $ Q_{\pm} \ce -i e^{i D \left(\frac{\omega}{2} \pm \Omega \right)}\erf{\left[i\frac{D}{2\sigma} + \sigma \left( \frac{\omega}{2} \pm \Omega \right)\right]}$ and $R_{\pm} \ce   \frac{D}{2}  \left( \frac{\omega}{2} \pm  \Omega\right) + i \left(1 + \frac{D^2}{4\sigma^{2}} \right) $.

\section{The effect of gravitational waves on vacuum correlations}
\label{corr}
In Sec.~\ref{detectorentanglementanalysis}, the dependence of the concurrence on the properties of gravitational waves and detectors was investigated, which quantifies the harvested entanglement in the final state of the detectors and is interpreted as a proxy for field entanglement. However, these detectors also harvest classical correlations from the vacuum. Thus, to quantify the total correlations harvested by a pair of detectors, interpreted analogously as a proxy for correlations between the region in which detectors interact, the correlations between local energy measurements (i.e., measurements of $\sigma_z$) can be computed. Such correlations are quantified by the correlation function~\cite{ martin-martinezSpacetimeStructureVacuum2016,smithDetectorsReferenceFrames2019}
\begin{align}
 \mbox{corr}\, \rho_{AB}& \ce \frac{\abs{X}^2 + \abs{C}^2}{P} +  \mathcal{O}\!\left(\lambda^4\right) = \Psi_{\mathcal{M}} + \Psi_{\rm GW}  + \mathcal{O}(\lambda^4) , \nn
 \label{definitionofcorr1}
\end{align}
where in the second equality the correlation function has been expressed as a sum of the Minkowski space and gravitational wave contributions to the correlation function, defined respectively as
\begin{align}
\Psi_{\mathcal{M}} &\ce  \frac{\abs{X_{\mathcal{M}} }^2 + \abs{C_{\mathcal{M}} }^2}{P} \nn,  \\
\Psi_{\rm GW} &\ce 2 \frac{ \Re[ X_{\rm GW} X_{\mathcal{M}}^* ] + \Re[C_{\rm GW} C_{\mathcal{M}}^* ]}{P}.
 \nn 
\end{align}

\begin{figure*}[t]
\subfloat[]{
\includegraphics[height = 2.2in]{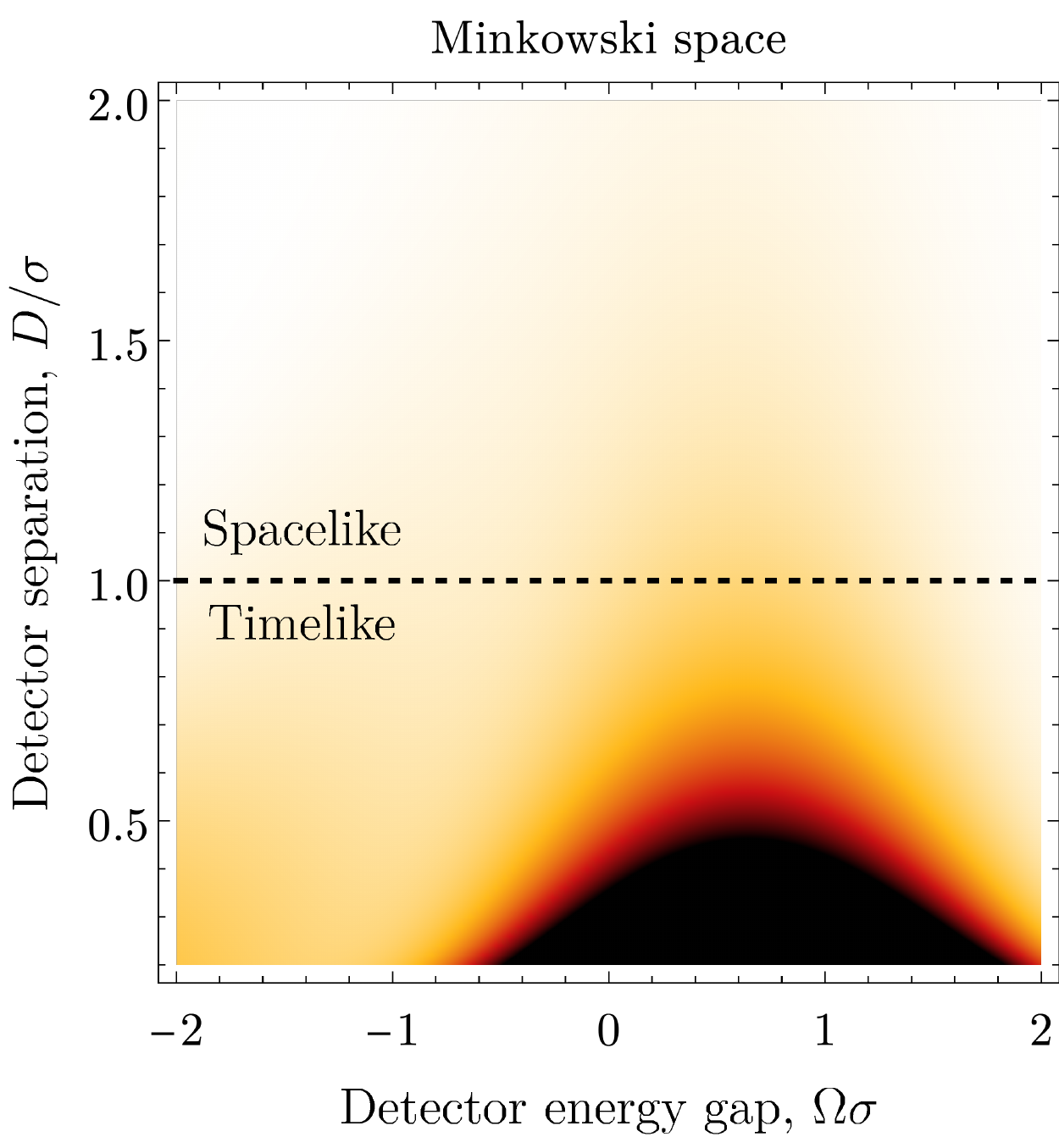} }
\subfloat[]{
\includegraphics[height = 2.2in]{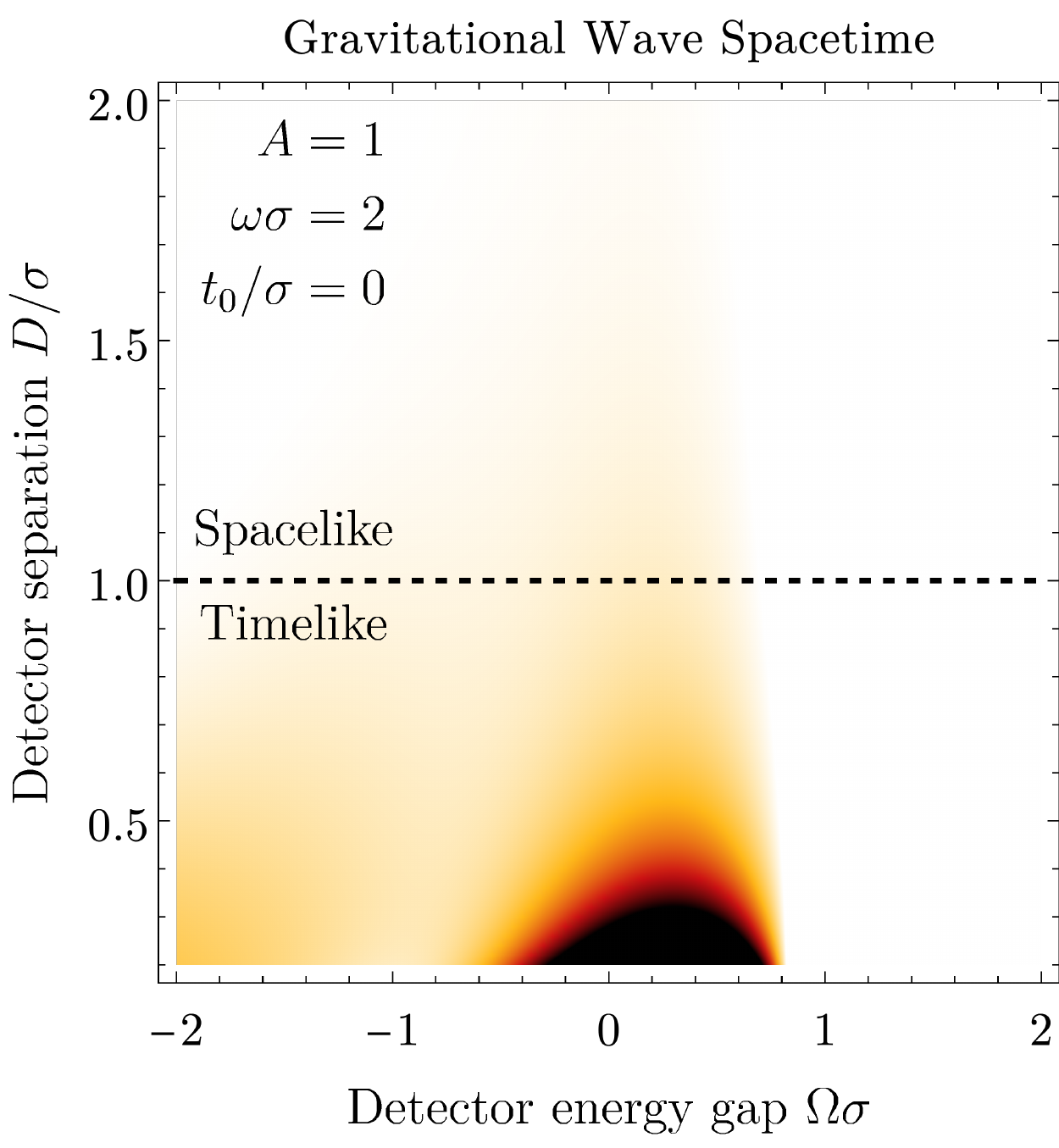} }
\subfloat[]{
\includegraphics[height = 2.2in]{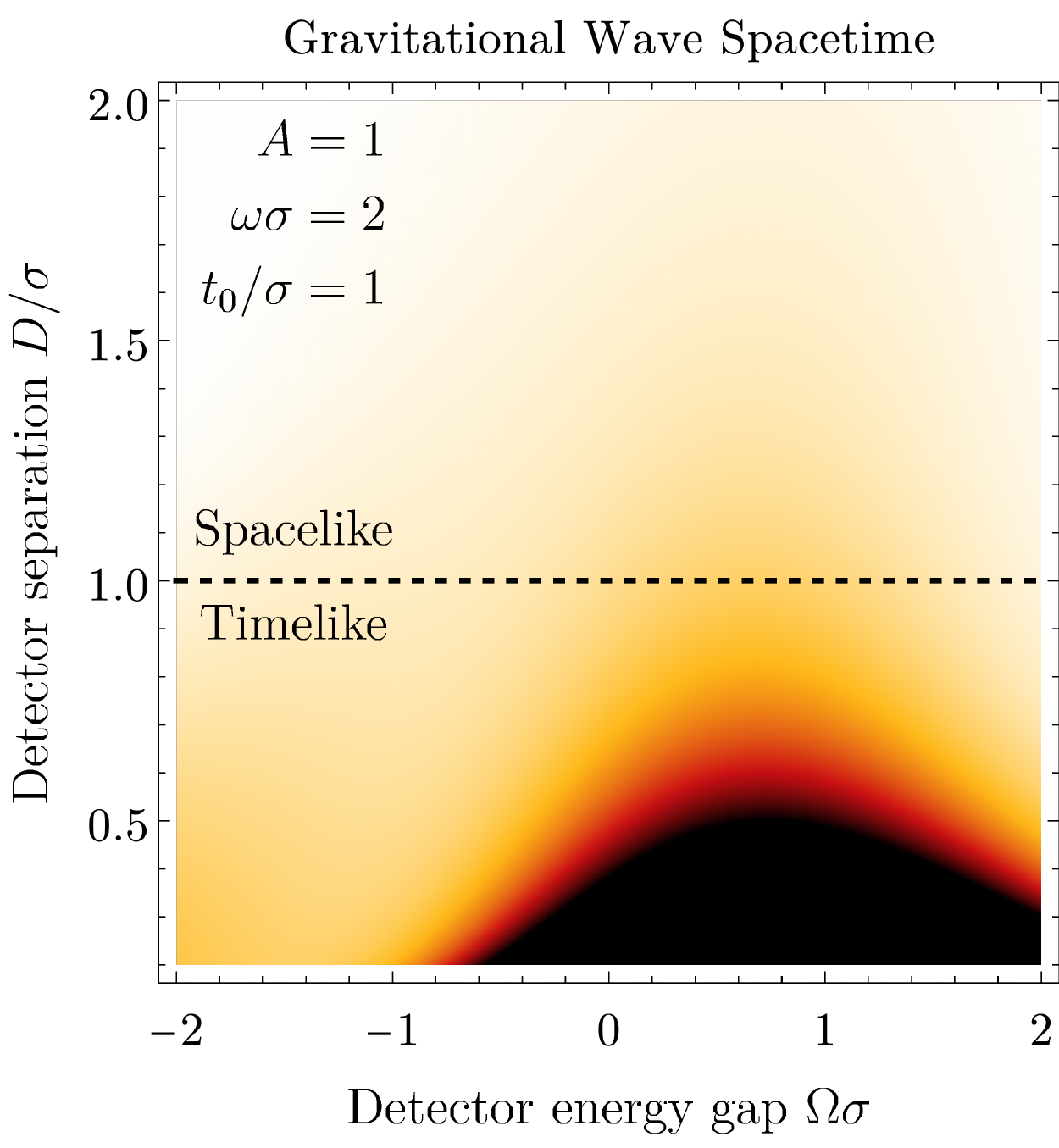} } 
\subfloat{
\includegraphics[height = 2.2in]{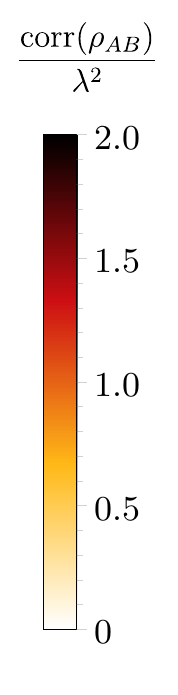} }

\caption{The correlation function ${\rm corr}(\rho_{AB})/\lambda^2$ is plotted as a function of the detectors' energy  $\Omega \sigma$ and the detectors' average proper separation $D / \sigma$ for detectors in (a) Minkowski space and detectors in a gravitational wave spacetime for (b) $t = 0$ and (c) $t \neq 0$. Analogous to the concurrence, we see that correlations between two detectors can be degraded or amplified depending on the value of $t_{0}$}
\label{fig:corr}
\end{figure*}

\begin{figure}[h]
\begin{center}
\includegraphics[height = 2.4in]{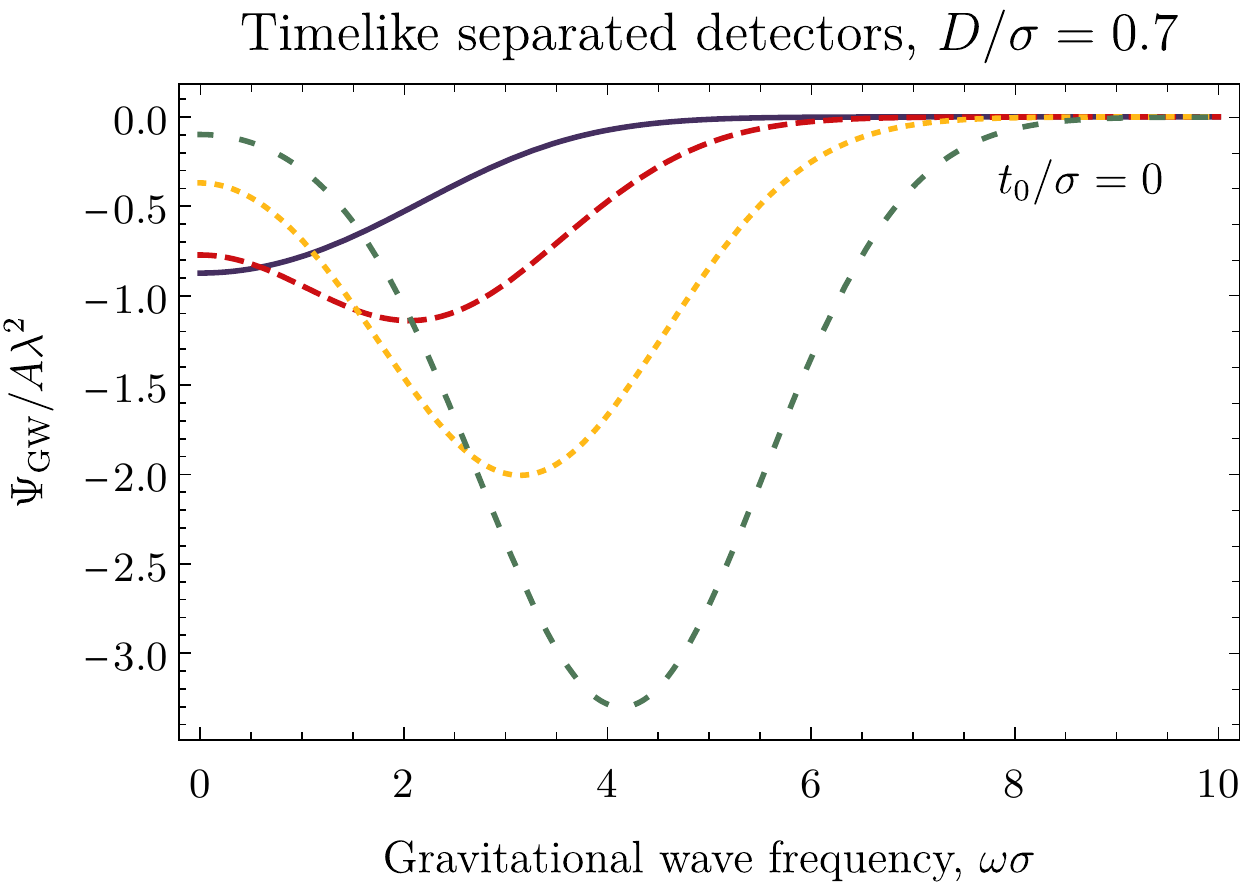} \quad
\includegraphics[height = 2.4in]{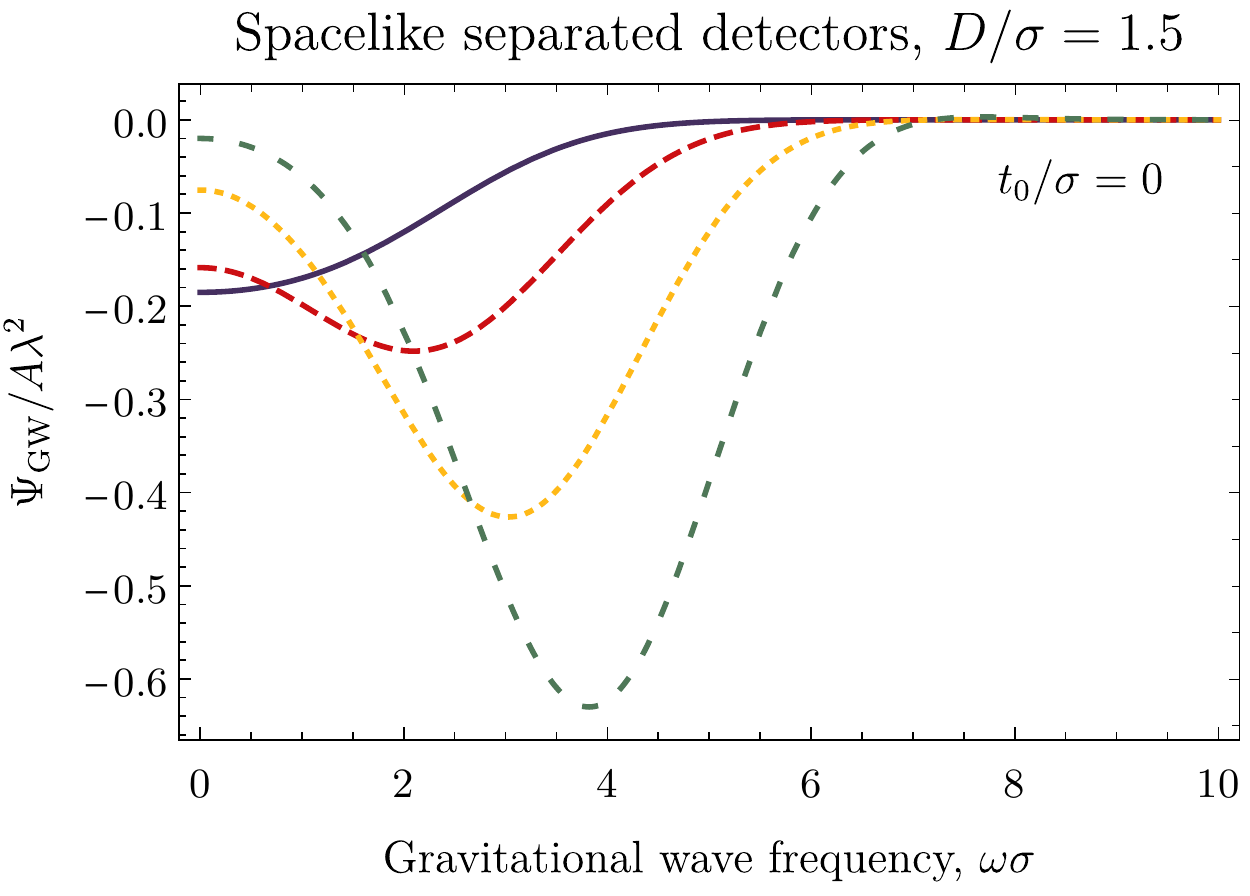} 
\end{center} 
\includegraphics[width = 3.25in]{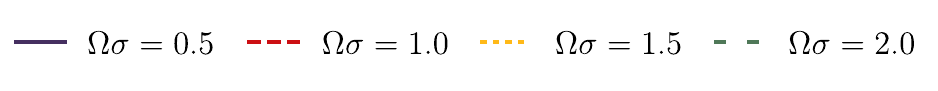} 
\caption{The gravitational wave contribution $\Psi_{\rm  GW}/A\lambda^2$ to the correlation function is plotted as a function of the gravitational wave frequency $\omega \sigma$ for both timelike (\emph{left}) and spacelike (\emph{right}) separated detectors for $t_{0}/\sigma = 0$ for different values of the detectors energy $\Omega \sigma$. Similar to $\Theta_{\rm GW}$, $\Psi_{\rm GW}$ is always negative, which implies that detector correlations are always degraded for  $t_0 = 0$.}

\label{fig:omegaVScorrt0}
\end{figure}

\begin{figure}[h]
\begin{center}
\includegraphics[height = 2.4in]{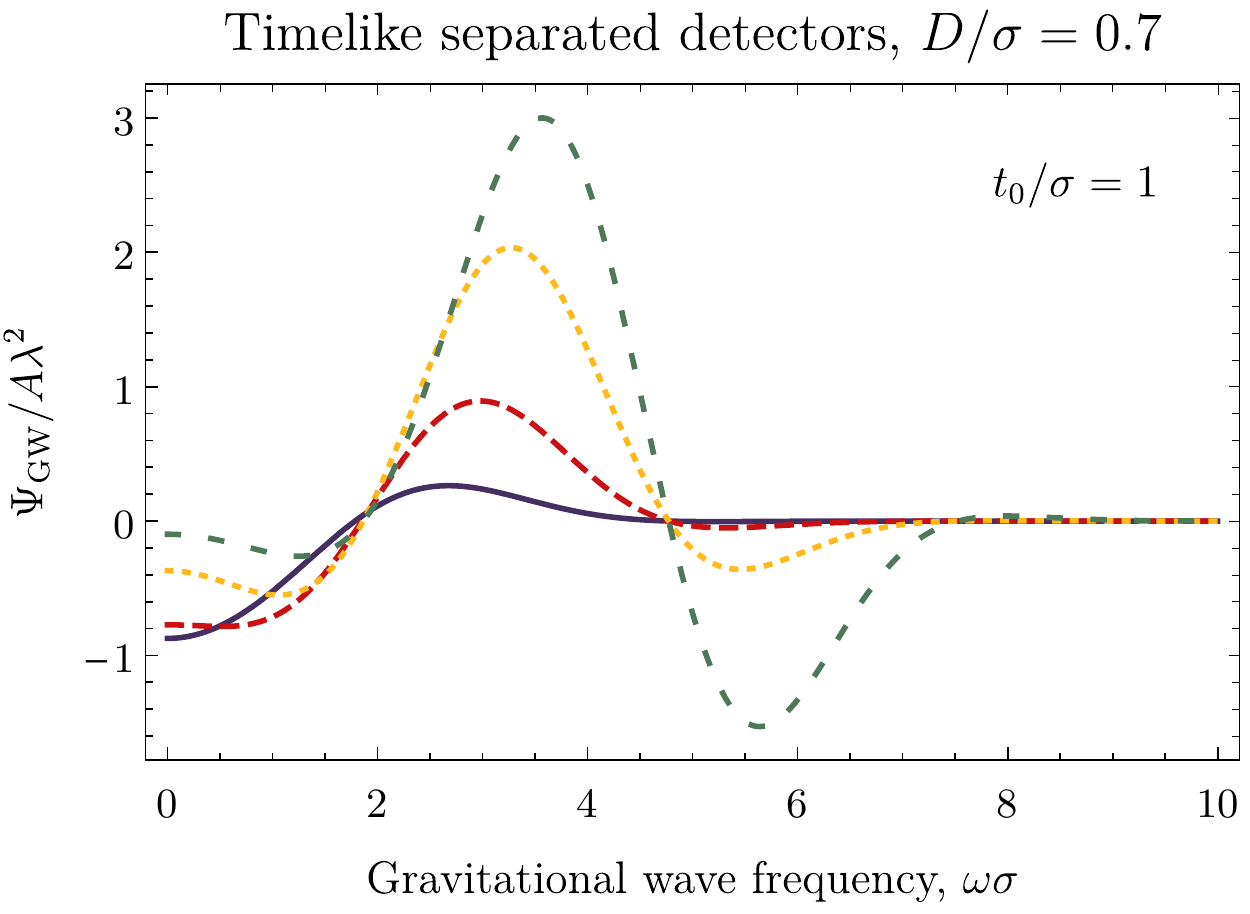} \quad
\includegraphics[height = 2.4in]{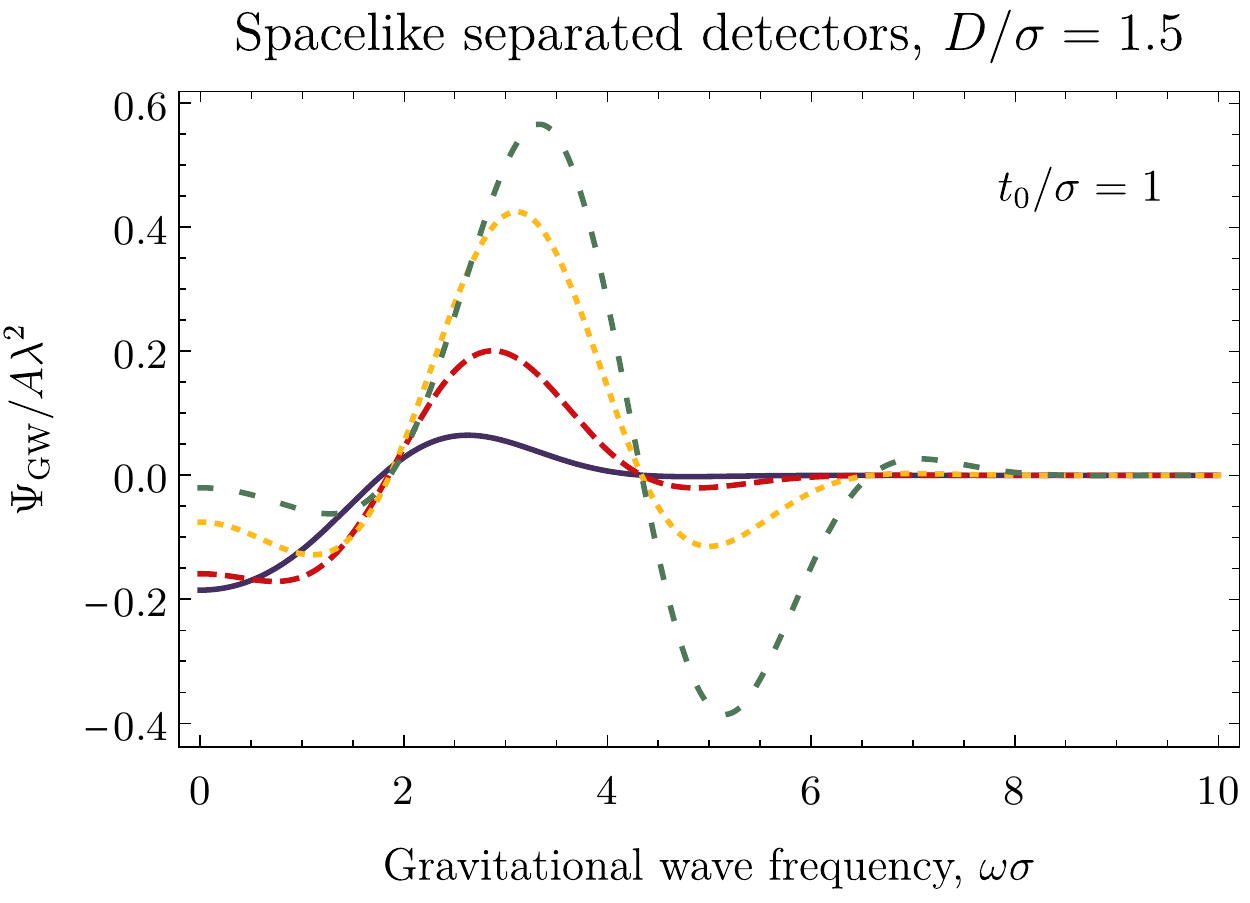} 
\end{center} 
\includegraphics[width = 3.25in]{PsiRowLegend.pdf} 
\caption{The gravitational wave contribution $\Psi_{\rm  GW}/A\lambda^2$ to the correlation function is plotted as a function of the gravitational wave frequency $\omega \sigma$ for both timelike (\emph{left}) and spacelike (\emph{right}) separated detectors for $t_{0}/\sigma =1$ for different values of the detectors energy $\Omega \sigma$. Similar to $\Theta_{\rm GW}$, $\Psi_{\rm GW}$ can be both positive and negative implying that a gravitational wave can amplify and degrade detector correlations.}
\label{fig:omegaVScorrt1}
\end{figure}

To examine the effect a gravitational wave has on the correlations harvested by the detectors, Fig.~\ref{fig:corr} compares correlations between detectors in Minkowski space with detectors in a gravitational wave spacetime, revealing similar behaviour as the concurrence depicted in Fig.~\ref{fig:Concurrence}. The gravitational wave contribution to the correlation function $\Psi_{\rm GW} $ is plotted in Figs.~\ref{fig:omegaVScorrt0} and \ref{fig:omegaVScorrt1} for $t_0=0$ and $t_0 \neq 0$, respectively. Similar to the concurrence, the correlation function exhibits a resonance around $\omega \approx 2 \Omega$ and oscillatory behaviour for nonzero $t_0$.

\end{document}